\documentclass[aip,apl,citeautoscript,twocolumn,reprint]{revtex4-1}

\usepackage{color}
\usepackage{graphicx}
\usepackage{dcolumn}
\usepackage{bm}
\usepackage{color}
\usepackage{tabularx}
\usepackage{array}
\usepackage{amsmath}
\usepackage{stmaryrd}  

\begin{document}

\title{\textcolor{black}{\uppercase{Quantum Engineering with Hybrid Magnonics Systems and Materials}}}

\author{\textit{(Alphabetical order)} \uppercase{D. D. Awschalom}, \textit{Member, IEEE}}
\affiliation{Pritzker School of Molecular Engineering, University of Chicago, Chicago, IL, USA}
\affiliation{Materials Science Division and Center for Molecular Engineering, Argonne National Laboratory, Lemont, IL, USA}
\author{\uppercase{C. H. R. Du}}
\affiliation{Department of Physics, University of California, San Diego, La Jolla, CA 92093, USA}
\author{\uppercase{R. He} }
\affiliation{Department of Electrical and Computer Engineering, Texas Tech University, Lubbock, TX 79409, USA}
\author{\uppercase{F. J. Heremans}, \textit{Member, IEEE}}
\affiliation{Pritzker School of Molecular Engineering, University of Chicago, Chicago, IL, USA}
\affiliation{Materials Science Division and Center for Molecular Engineering, Argonne National Laboratory, Lemont, IL, USA}
\author{\uppercase{A. Hoffmann}, \textit{Fellow, IEEE}}
\affiliation{Materials Research Laboratory and Department of Materials Science and Engineering, University of Illinois at Urbana-Champaign, Urbana, IL 61801, USA}
\author{\uppercase{J. T. Hou}, \textit{Member, IEEE}}
\affiliation{Electrical Engineering and Computer Science, Massachusetts Institute of Technology, Cambridge, MA, USA}
\author{\uppercase{H. Kurebayashi}}
\affiliation{London Centre for Nanotechnology, University College London, 17-19 Gordon Street, London, WCH1 0AH, UK}
\author{\uppercase{Y. Li}, \textit{Member, IEEE}}
\affiliation{Materials Science Division, Argonne National Laboratory, Lemont, IL, USA}
\author{\uppercase{L. Liu}, \textit{Member, IEEE}}
\affiliation{Electrical Engineering and Computer Science, Massachusetts Institute of Technology, Cambridge, MA, USA}
\author{\uppercase{V. Novosad}, \textit{Member, IEEE}}
\affiliation{Materials Science Division, Argonne National Laboratory, Lemont, IL, USA}
\author{\uppercase{J. Sklenar}, \textit{Member, IEEE}}
\affiliation{Department of Physics and Astronomy, Wayne State University, Detroit, MI 48201, USA}
\author{\uppercase{S. E. Sullivan}}
\affiliation{Materials Science Division and Center for Molecular Engineering, Argonne National Laboratory, Lemont, IL, USA}
\author{\uppercase{D. Sun}}
\affiliation{Department of Physics, North Carolina State University, Raleigh, NC 27695, USA}
\author{\uppercase{H. Tang}, \textit{Member, IEEE}}
\affiliation{Department of Electrical Engineering, Yale University, New Haven, CT, USA}
\author{\uppercase{V. Tiberkevich}, \textit{Member, IEEE}}
\affiliation{Department of Physics, Oakland University, MI 48309, USA}
\author{\uppercase{C. Trevillian}, \textit{Member, IEEE}}
\affiliation{Department of Physics, Oakland University, MI 48309, USA}
\author{\uppercase{A. W. Tsen}}
\affiliation{Institute for Quantum Computing and Department of Chemistry, University of Waterloo, Waterloo, ON, Canada}
\author{\uppercase{L. R. Weiss}}
\affiliation{Pritzker School of Molecular Engineering, University of Chicago, Chicago, IL, USA}
\author{\uppercase{W. Zhang}, \textit{Member, IEEE}}
\affiliation{Department of Physics, Oakland University, MI 48309, USA}
\author{\uppercase{X. Zhang}}
\affiliation{Center for Nanoscale Materials, Argonne National Laboratory, Lemont, IL, USA}
\author{\uppercase{L. Zhao}}
\affiliation{Department of Physics, University of Michigan, Ann Arbor, MI 48109, USA}
\author{\uppercase{C. W. Zollitsch}}
\affiliation{London Centre for Nanotechnology, University College London, 17-19 Gordon Street, London, WCH1 0AH, UK}

\date{\today}

\begin{abstract}
\textcolor{black}{Quantum technology has made tremendous strides over the past two decades with remarkable advances in materials engineering, circuit design and dynamic operation. In particular, the integration of different quantum modules has benefited from hybrid quantum systems, which provide an important pathway for harnessing the different natural advantages of complementary quantum systems and for engineering new functionalities. This review focuses on the current frontiers with respect to utilizing magnetic excitatons or magnons for novel quantum functionality. Magnons are the fundamental excitations of magnetically ordered solid-state materials and provide great tunability and flexibility for interacting with various quantum modules for integration in diverse quantum systems. The concomitant rich variety of physics and material selections enable exploration of novel quantum phenomena in materials science and engineering. In addition, the relative ease of generating strong coupling and forming hybrid dynamic systems with other excitations makes hybrid magnonics a unique platform for quantum engineering. We start our discussion with circuit-based hybrid magnonic systems, which are coupled with microwave photons and acoustic phonons. Subsequently, we are focusing on the recent progress of magnon-magnon coupling within confined magnetic systems. Next we highlight new opportunities for \textcolor{black}{understanding} the interactions between magnons and nitrogen-vacancy centers for quantum sensing and implementing quantum interconnects. Lastly, we focus on the spin excitations and magnon spectra of novel quantum materials investigated with advanced optical characterization.}

\end{abstract}

\maketitle

\section{Introduction}
\label{sec:introduction}


\textcolor{black}{Quantum technology combines fundamental quantum physics and information theory with an overarching goal to develop a new generation of computing, sensing, and communication architectures based on quantum coherent transfer and storage of information. Spurred by the discovery of quantum properties in novel materials and structures, a variety of dynamic systems including photons, acoustic excitations, and spins have been cultivated in diverse platforms such as superconducting circuits, nanomechanical devices, surface acoustic waves, and individual electrons and ions. Many of these systems have been employed for implementing artificial two-level systems as qubits, enabling their coherent interaction as well as manipulating their states. For nascent quantum technologies to reach maturity, a key step is the development of scalable quantum building blocks: from quantum interconnects and transducers, to sensors at the single quasiparticle level. The development of scalable architectures for quantum technologies not only poses challenges in understanding the coupling between disparate quantum systems, but also presents technical and engineering challenges associated with developing chip-scale quantum technology.}

\textcolor{black}{A rapidly growing subfield of quantum engineering is associated with magnons. Magnons, or the quanta of spin waves, are the collective excitation of exchange coupled spins hosted in magnetic materials. Similar to electromagnetic and acoustic waves, spin waves can propagate and interfere, meaning that they can deliver phase information for coherent information processing. Due to the high spin density in magnetic materials compared with individual spins, large magnetic dipolar coupling strengths in the sub-gigahertz regime can be easily achieved between magnons and microwave photons, which means fast operation and transduction before decoherence. In addition, their frequency can be readily manipulated by magnetic fields, and thus either spatially or temporally varying magnetic fields enable straightforward adiabatic modifications. More importantly, magnons can provide a wide range of interactions. For example, magnons have been demonstrated to mediate coupling between microwave and optical photons via magneto-optic Faraday effects. Magnons also can interact strongly with phonons due to magnetoelastic coupling. The \textcolor{black}{inherent} strongly nonlinear dynamics of magnons makes them also very susceptible to interactions with other magnons. Meanwhile, the interaction of magnons and spins such as nitrogen-vacancy (NV) centers \textcolor{black}{in diamond} brings new ideas for quantum sensing of magnons and coherent manipulation of spin qubits. From a materials perspective, the complex interactions between electric charge currents and magnetization dynamics, which are key to modern spintronics concepts,\cite{HoffmannPRAppl2015} provides additional pathways of magnon generation and evolution.} 
%
\textcolor{black}{
While most current research efforts are focused on quasi-classical magnon coupling to other quantum systems, recent studies of non-classical magnon states \cite{KamraAPL117_2020} and the discovery of atomically thin magnetic materials show promise for observing and utilizing new genuinely quantum effects in magnon dynamics.
}
\textcolor{black}{In this review, we explore the evolving boundary between quasi-classical behavior and quantum interactions enabled by the rich physics of magnonics. We delineate the ongoing materials and device engineering efforts that are critical to enabling the next generation of quantum technologies based on magnons, which } \textcolor{black}{will significantly broaden the scope of fundamental quantum research and augment the functionality in quantum information processing.} 


While many similar interactions may also be possible with individual spins, the collective dynamics of the magnetically ordered systems provides distinct advantages with easily detectable signals and high coupling efficiencies while maintaining reasonable quality factors up to $\sim10^5$.  At the same time, due to the aforementioned beneficial scaling properties, hybrid magnon systems are very well suited to on-chip integration,\cite{LiJAP2020} \textcolor{black}{as will be} discussed in greater detail in Sec.~\ref{sec:Hybrid Magnonics Systems}.  In addition, their dynamics can provide pronounced non-reciprocities, which are beneficial for the unidirectional flow of quantum information.

In Section III, we discuss how layered antiferromagnets are exemplary materials for the study of magnon-magnon interactions.  Both synthetic antiferromagnets and van der Waals magnets have an antiferromagnetic interlayer exchange interaction that enables acoustic and optical magnons.  Unique to these materials is that the interaction between acoustic and optical magnons (or even amongst optical and acoustic magnons) can be continuously tuned via field orientation\cite{MacNeill2019,Sud2020}, wave number\cite{Shiota2020}, or magnetic damping\cite{sklenar2020self}.  As a consequence, the magnon energy spectrum can be manipulated \textcolor{black}{via the tunable magnon-magnon interaction strength.} 
The examples we highlight have weaker antiferromagnetic interactions that enable optical magnons to exist at GHz frequencies\textcolor{black}{, compatible with the characteristic frequency range of existing qubit technologies}. The strategies used in these materials can be potentially deployed in antiferromagnets with characteristic THz magnons. 


In Section \ref{sec:sensing}, we explore optically addressable solid-state qubits derived from atomic defects in semiconductors and their coupling to magnons in magnetic materials. In particular, color centers like the NV center in diamond \cite{Heremans2016} combine the ability to optically detect and manipulate the qubit spin state with high sensitivity to proximal magnetic fields, making them excellent sensors for probing magnons \cite{Du2017} and potential quantum transducers for upconverting GHz qubit excitations to optical photons \cite{Hensen2015}. In addition to sensing with solid-state defects, we discuss the important materials engineering considerations for on-chip quantum hybrid systems based on magnons and solid-state defects, from theoretical to materials considerations for both the defect material itself and the magnetic materials that host spin wave excitations.

\textcolor{black}{In Section \ref{sec:materials}, we discuss novel magnetic excitations and effects in quantum materials and their detection via optical spectroscopy. In particular, we highlight three families of magnetic quantum materials that have drawn extensive research interest, including two-dimensional (2D) magnetic atomic crystals, strong spin-orbit coupled (SOC) $J_{eff}=1/2$ magnetism in 5$d$ transition metal oxides, and assembled molecular chiral spin and magnon systems. We show the recent development of high-sensitivity, symmetry-resolved magneto-Raman spectroscopy in detecting magnetic excitations in 2D magnets and SOC magnets and the progress of ultra-sensitive magneto-optical Kerr effect (MOKE) of Sagnac interferometers in probing the chirality induced spin selectivity (CISS) effect in molecular spin and magnon systems. \textcolor{black}{For each highlighted development,} we further comment on the prospects of controlling the magnetic properties and integrating them with spintronic devices.}\\

\section{Directions for ``On-chip" quantum platforms with Hybrid Magnonics}
\label{sec:Hybrid Magnonics Systems}

\subsection{\textcolor{black}{Magnon-Photon Coupling and Superconducting resonators}}

Magnon-photon hybrid dynamic systems have recently attracted great attentions due to their rich physics, strong coupling and convenience for microwave engineering \cite{HueblPRL2013,TabuchiPRL2014,ZhangPRL2014,2014_PRApplied_Goryachev,BhoiJAP2014,2015_PRL_Bai}. Specifically, the emerging field of quantum magnonics aims to realize the potential offered by magnons for quantum information \cite{LachanceQuirionAPEx2019}, with the recent demonstration of coherent coupling between a single magnon and a superconducting qubit and single-shot readout of magnon numbers \cite{TabuchiScience2015,LachanceScienceAdvan2017,LachanceQuirionScience2020}. As collective excitations of exchange coupled spins, magnons enable large coupling strengths to microwave photons due to the high spin density in magnetic materials. Furthermore, magnons exhibit excellent frequency tunability and can interact with different excitations such as optical photons, phonons and spins, making them suitable for exploring fundamental physics in coherent dynamics as well as constructing highly tunable quantum transducers.

On-chip implementation and integration of quantum magnonic systems are highly desirable for circuit-based applications and building up complex networks of coupled magnonic systems \cite{BabakPRB2018,WangPRApplied2018}. One direct approach is to integrate magnetic devices with coplanar superconducting resonators \cite{HueblPRL2013,MorrisSREP2017,McKenziePRB2019,LiPRL2019,HouPRL2019,MandalarXiv2020}, which can carry long-coherence microwave photons and couple with superconducting qubits for building up circuit quantum electrodynamics (cQED).

\begin{figure*}[htb]
 \centering
 \includegraphics[width=6.0 in]{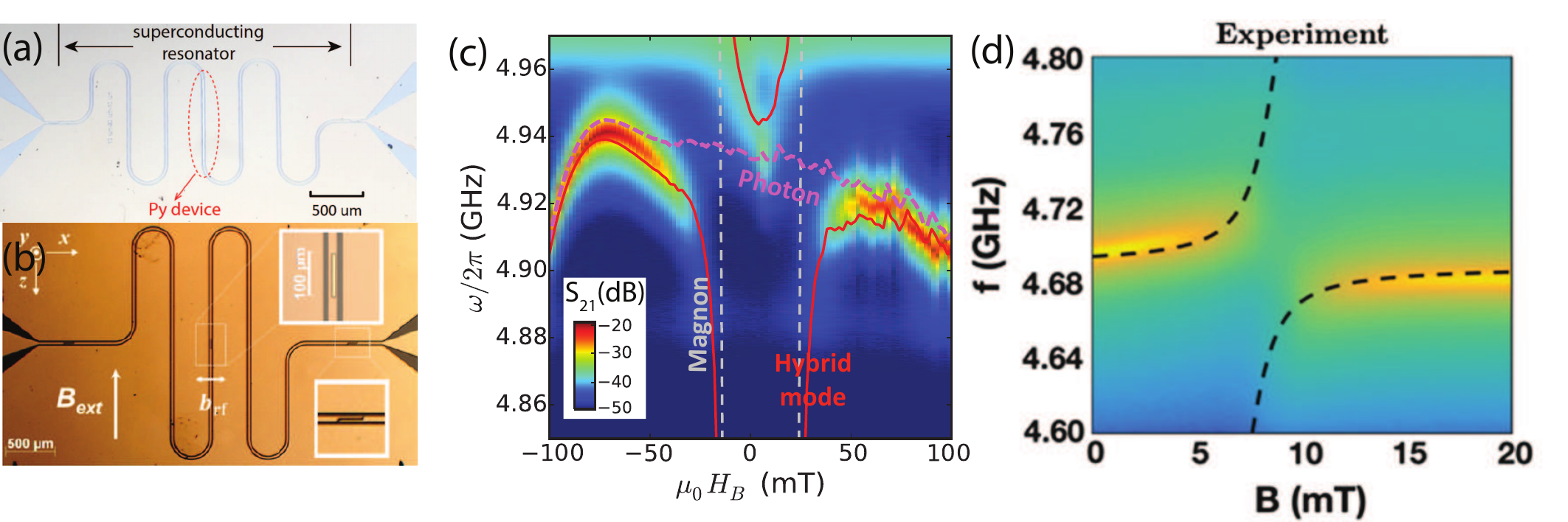}
 \caption{Strong magnon-photon coupling between Ni$_{80}$Fe$_{20}$ (permalloy, Py) devices and superconducting resonators. (a-b) Optical microscope images of half-wavelength superconducting resonators with Ni$_{80}$Fe$_{20}$ devices fabricated at the center, with (a) fabricated from NbN thin film and (b) from Nb film. (c-d) Mode anti-crossing spectra between Ni$_{80}$Fe$_{20}$ magnon modes and the photon modes of superconducting resonators. (a) and (c) are adapted from Ref.~\onlinecite{LiPRL2019}; (b) and (d) are adapted from Ref.~\onlinecite{HouPRL2019}.}
 \label{fig_sc01}
\end{figure*}

The coupling strength between magnons and microwave photons can be expressed as:
\begin{equation}\label{eq_sc01}
g= \gamma \sqrt{\frac{N\mu_0\hbar\omega_p}{4V_c}}.
\end{equation}
where $\omega_p$ denotes the photon frequency of the microwave cavity, $N$ denotes the total number of spins and $V_c$ denotes the effective volume of the microwave cavity. Note that $N$ can be converted to the effect magnetic volume $V_M$ by $N\mu_B=M_sV_M$ where $M_s$ is the magnetization and $\mu_B=\gamma\hbar$ is the single Bohr magneton momentum. The factor of $1/4$ comes from two effects: i) the magnetic field accounts for only half of the total energy in the microwave cavity; and, ii) linear polarized microwave fields need to be decomposed into two circularly polarized components and only one component couples to magnon excitations. From Eqs.~(\ref{eq_sc01}), in order to increase $g$ with a limited magnetic device volume $V_M$ (or $N$), it is important to have a small $V_c$, which leads to a large coupling strength per Bohr magneton defined as $g_0 = g / \sqrt{N}$.


Recently, Li \textit{et al.}\ \cite{LiPRL2019} and Hou \textit{et al.}\ \cite{HouPRL2019} have demonstrated strong magnon-photon coupling between permalloy (Ni$_{80}$Fe$_{20}$, Py) thin-film devices and coplanar superconducting resonators. As shown in Figs.~\ref{fig_sc01}(a) and (b), in both works a Py device was fabricated in the middle of a superconducting resonator, which is a coplanar waveguide with two capacitive couplers on the two side defining the half wavelength and the resonant frequency. With the microwave transmission data shown in Figs.~\ref{fig_sc01}(c) and (d), an in-plane magnetic field was applied to modify the magnon frequency of the Py device. Clear avoided crossings are observed when the magnon mode intersects with the resonator photon mode. Li \textit{et al.}\ \cite{LiPRL2019} have achieved a coupling strength of 152~MHz with a 900$-\mu$m$\times 14-$ $\mu$m$\times 30-$nm Py stripe and Hou \textit{et al.}\ \cite{HouPRL2019} have reported a coupling strength of 171~MHz for a 2000$-\mu$m$\times 8-$ $\mu$m$\times 50-$nm stripe. In addition, both reports have shown that the coupling strength scales with $\sqrt{V_M}$ by changing the thickness or the length of the Py device, agreeing with the prediction of Eq.~(\ref{eq_sc01}).

The key for reaching sub-gigahertz coupling strength with small Py devices, as compared with the macroscopic yttrium iron garnet (YIG) crystals \cite{TabuchiPRL2014,ZhangPRL2014} is that the effective volume of the coplanar microwave resonator is much smaller than 3D cavities. Li \textit{et al.}\ have estimated $V_c=0.0051$ mm$^3$, leading to efficient coupling of $g_0/2\pi=26.7$~Hz. Hou \textit{et al.}\ have obtained a similar sensitivity of $g_0/2\pi=18$~Hz. In addition, a high quality factor of the superconducting resonator also ensures large cooperativity which is important for coherent operation and transduction. At a temperature of 1.5 K, the two works show quality factors of $Q=7800$ from Ref.~\onlinecite{LiPRL2019} and 1520 from Ref.~\onlinecite{HouPRL2019} for their superconducting resonator. Although the quality factor will be reduced when the Py device is fabricated, a large cooperativity of $C=68$ \cite{LiPRL2019} and 160 \cite{HouPRL2019} can still be achieved and the strong coupling regime is obtained.

\begin{figure}[htb]
 \centering
 \includegraphics[width=3.0 in]{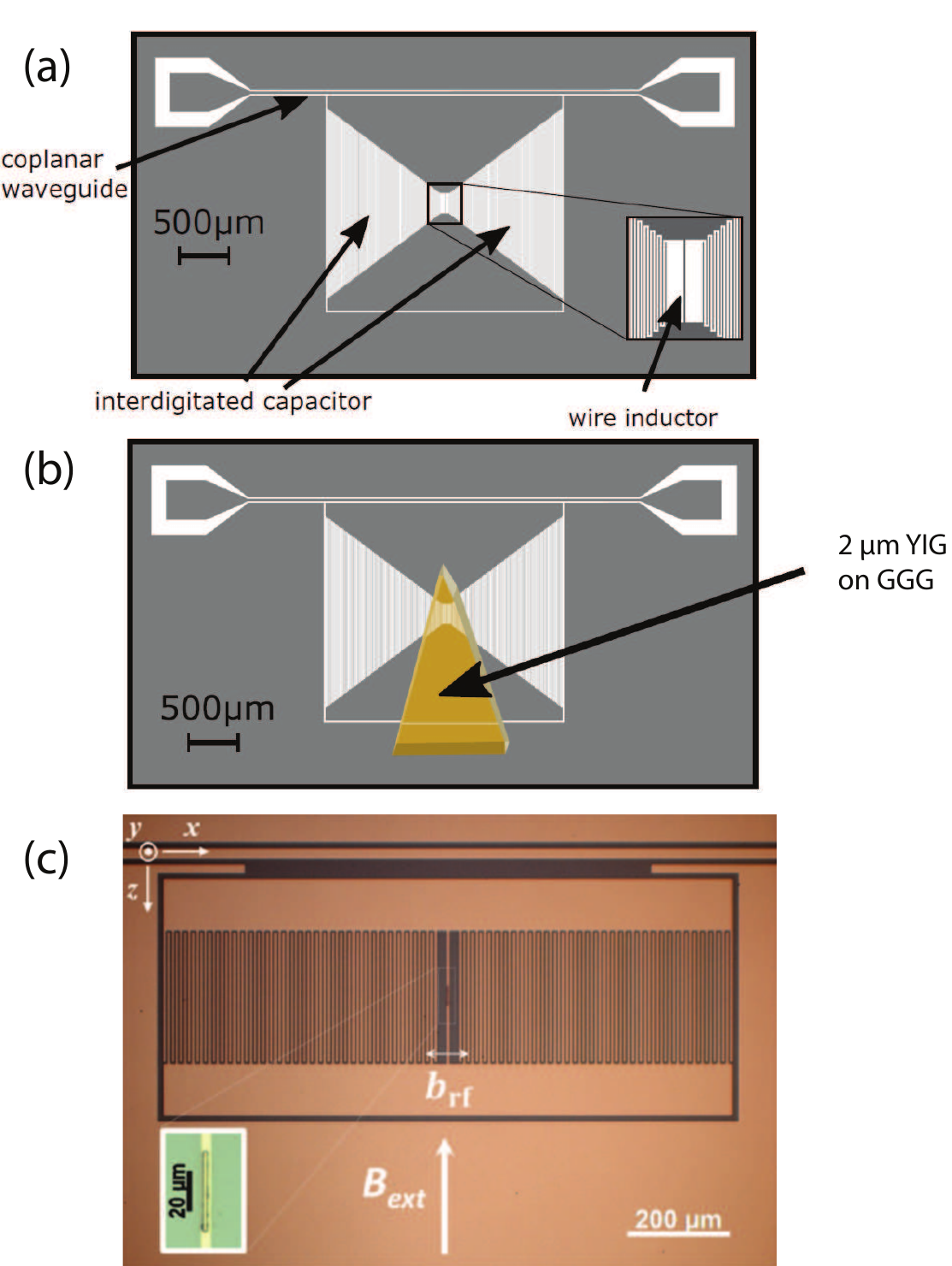}
 \caption{Lumped element resonator (LER) design. (a-b) LER resonator coupled to a flip-chip YIG film grown on a GGG substrate, (a) without and (b) with a YIG film. (c) LER coupled to a Py device fabricated at the center inductive wire. (a) and (b) are adapted from Ref.~\onlinecite{McKenziePRB2019}; (c) is adapted from Ref.~\onlinecite{HouPRL2019}.}
 \label{fig_sc02}
\end{figure}

The beauty of coplanar superconducting resonators is that they can be designed with great flexibility while still maintaining a high quality factor. To further reduce the effective volume, a lumped element resonator (LER) design has been used by McKenzie-Sell \textit{et al.}\ \cite{McKenziePRB2019} and Hou \textit{et al.}\ \cite{HouPRL2019}. As illustrated in Fig.~\ref{fig_sc02}(a), the LER consists of a small wire inductor and a large interdigitated capacitor, which minimize the inductive volume for large magnetic coupling efficiency while maintaining a balanced resonant frequency with a large capacitance. The inverse design with minimized capacitor and maximized inductor has been also used for increasing the electrical coupling efficiency \cite{HanPRL2016}. In Fig.~\ref{fig_sc02}(b), by covering the middle wire inductor with a small piece of 2-$\mu$m-thick YIG film, a magnon-photon coupling strength of 300~MHz has been achieved. In Fig.~\ref{fig_sc02}(c), a coupling strength of 74.5~MHz has been achieved with a 40$-\mu$m$\times 2$ $-\mu$m$\times 10-$nm Py device. The latter corresponds to a sensitivity of $g_0/2\pi=263$~Hz. This value agrees with a prior report of LER \cite{EichlerPRL2017}, with $g_0/2\pi=150$~Hz. In Ref.~\onlinecite{EichlerPRL2017} the quality factor of the loaded LER can reach $Q=1400$, which is sufficient for a clear observation of avoided crossing between the magnon and photon modes. 

For all-on-chip integration of magnon-photon hybrid systems with superconducting resonator, in addition to the resonator design, the search for low-damping and fabrication-friendly magnetic materials is also crucial. While Py is a classical ferromagnet with convenient deposition and fabrication, its relatively large Gilbert damping is not optimal for long-coherence magnon-photon interaction. For low-damping YIG thin films, one issue is that they are typically grown on gadolinium gallium garnet (GGG) substrates, which possess a complex magnetic behavior at cryogenic temperature \cite{MihalceanuPRB2018,KosenAPLMaterials2019} and will increase the loss of excitations if the superconducting circuits are fabricated on GGG substrate. To address this issue, free-standing YIG thin films \cite{SeoAPL2017,HeyrothPRApplied2019} or YIG films grown on Si substrate \cite{FanPRApplied2020} may provide a solution. Conversely, since flip-chip arrangements of YIG/GGG sample have been used to achieve strong coupling with superconducting resonators, it is also possible to separate the fabrication of superconducting resonators and the YIG magnonic structures and then mount the two systems together by a flip-chip technique, which has been applied in quantum acoustics \cite{SatzingerNature2018}.

\subsection{\textcolor{black}{Magnetomechanics and Magnon-Phonon Coupled Devices}}


The study of magnon-phonon interaction dates back to the 1950s and 1960s, when the strong coupling between spin waves (magnons) and acoustic waves (phonons), as well as the resulting magnetoelastic waves (hybrid magnon-phonon states) have been investigated in both theory and experiments \cite{1958_PR_Kittel,1960_JAP_Schlomann,1963_JAP_Eshbach_ME,1964_JAP_Schlomann_ME,1965_ProcIEEE_Damon,1967_APL_Auld,1967_JAP_Comstock,1969_JAP_Rezende}. In recent years, with the increasing interests in utilizing phonons as an information carrier for coherent and quantum information processing \cite{2008_Nature_Li,2010_Science_Weis,2011_Nature_Chan,2011_Nature_Teufel}, magnon-phonon coupled systems have re-emerged as a promising platform for hybrid magnonics. Phonons exhibit very long lifetimes in solid state platform such as silicon, quartz, diamond, aluminum nitride, lithium niobate, etc., which satisfy the requirements of a large variety of coherent applications. However, the frequency of phonon modes, although they can be conveniently tailored by proper geometry engineering, are usually fixed. On the other hand, as a coherent information carrier, magnons possess superior tunability but suffer from their finite lifetimes. Therefore, there are emerging needs for hybridizing magnons and phonons to combine their respective advantages for coherent and quantum information processing.

The most prominent underlying mechanism for magnon-phonon coupling in magnetic media such as YIG is the magneostrictive effect, which links magnetization with strain in magnetic materials. Although magnons represent small, typically weak perturbations of the magnetization, the associated dynamical spin precession can still efficiently couple with the long-lived phonon modes thanks to the excellent mechanical properties of the material. In particular, as in magnon-photon coupling, the magnon-phonon coupling is greatly enhanced by the large spin density \cite{2016_SciAdv_Zhang}. 

In general, the interaction between magnons and phonons, which is referred to as magnetomechanical coupling, can take two different forms, considering that the magnon modes usually fall into the GHz frequency range while the phonon modes can be tailored in a broad frequency range from kHz or even lower to GHz. In the first scheme, the interacting magnon mode and phonon mode have identical frequencies and the interaction is described by a beam-splitter type Hamiltonian $H_\mathrm{int}=\hbar G(mb^\dagger+m^\dagger b)$, where $m$ and $m^\dagger$ ($b$ and $b^\dagger$) are the creation and annihilation operator of the magnon (phonon) mode, respectively, and $G$ is the coupling strength that is enhanced by the large spin number. In the other scheme, the phonon mode is at a much lower frequency compared with the magnon mode, and they interact with each other with the assistance of a parametric drive, which can be described by a radiation-pressure type Hamiltonian $H_\mathrm{int}=\hbar G m^\dagger m(b+b^\dagger)$. In this case, the parametric magnomechanical interaction is not only enhanced by the large spin density but can also be boosted by the strong drive power.

The magnetomechanical devices can take various forms. In the most commonly used YIG sphere resonators, radiation-pressure type magnetomechanical interaction is experimentally demonstrated by Zhang and co-workers in 2016 \cite{2016_SciAdv_Zhang}. Benchmark coherent phenomena, including magnetomechanically induced transparency/absorption (MMIT/MMIA), parametric amplification, and phonon lasing has been observed, revealing the great applied potential of cavity magnetomechanics. Compared with optomechanical or electromechanical systems, magnomechanical systems benefit from the intrinsic magnon properties and exhibit unprecedented tunability. Moreover, such magnetomechanical systems support mode hybridization not only between magnons and phonons but also between magnons and microwave photons, which further enriches the system dynamics. Since then, various novel physical phenomena or applications based on magnetomechanical interactions have been proposed or experimentally studied, including magnon blockade \cite{2020_AnnPhys_Wang}, phonon-mediated magnon entanglement \cite{2019_NJP_Li}, magnetomechanical squeezing \cite{2019_PRA_Li_squeeze}, magnon-assisted ground state cooling of phonon state \cite{2020_Josa_Ding_Phononlaser}, etc.

Beam-splitter type magnetomechanical interactions are usually observed in planar structures with YIG thin films grown on GGG substrates. Compared with the mechanical modes in spherical geometries which are typically in the MHz range, acoustic modes in planar structures can be excited at GHz frequencies and therefore can directly coupled with magnons. Among many acoustic excitation forms, bulk acoustic waves (BAWs) are often used for hybridizing with magnons. BAW is a type of volume excitation with the mechanical displacements distributed in the whole substrate including both the magnetic YIG layer and the non-magnetic GGG layer. Considering that magnons only reside in the top YIG layer, their interaction is far from optimal. However, because of the excellent material property of single crystalline YIG and GGG, the dissipation of BAWs is extremely low. As a result, the magnetomechanical coupling can still exceed the system dissipations and reach the strong coupling regime \cite{2020_PRB_An_magnonBAW,2020_APL_Alekseev}. Thanks to the non-localized nature of BAWs, the magnon-phonon strong coupling can be utilized to mediate long-range interactions between magnon modes \cite{2020_PRB_An_magnonBAW}. 

In addition to BAWs, surface acoustic waves (SAWs) have also been adopted for coupling with magnons \cite{2017_JAP_Li_magnonSAW,2019_PRAppl_Duquesne_magnonSAW,2018_PRB_Xu_magnonSAW}. Unlike BAW-based magnetomechanical interactions, SAW-based magnon-phonon interactions have been carried out on more diverse material platforms. Although SAWs normally suffer from lower frequencies and higher losses \cite{2008_IFCS_Fattinger}, and usually requires heterogeneous integration with piezoelectric materials such as lithium niobate or gallium nitride, they exhibit better mode matching and consequently improved coupling with magnons because their mechanical displacements are mainly localized on the surface of the device, which is distinctively different from BAWs. In addition, the unique behaviors of SAW also enabled the experimental observation of a novel coupling mechanism for magnon-phonon interactions that has been predicted decades ago: the magneto-rotation coupling \cite{2020_SciAdv_Xu_magnonSAW}, which may lead to novel physical phenomena and applications.

The interaction between magnons and phonons has led to many novel functionalities. On the one hand, magnon lifetimes are usually on the order of 100 nanosecond \cite{1961_PR_Spencer,ZhangPRL2014} while phonon lifetimes can be several orders of magnitude longer \cite{2016_SciAdv_Zhang}, thus magnon systems benefit from their hybridization with phonons resulting in significantly improved coherence. On the other hand, phonons also gain new properties that are intrinsically missing in the mechanical degree of freedom. For instance, through hybridization with magnon modes, phonons can inherit their magnetic characteristics and carry spins \cite{2018_NPhys_Holanda_phononSpin} or support unidirectional propagation \cite{2020_PRL_Zhang_unidirectPump,2020_SciAdv_Xu_magnonSAW}.

\begin{figure*}[htb]
 \centering
 \includegraphics[width=5.5 in]{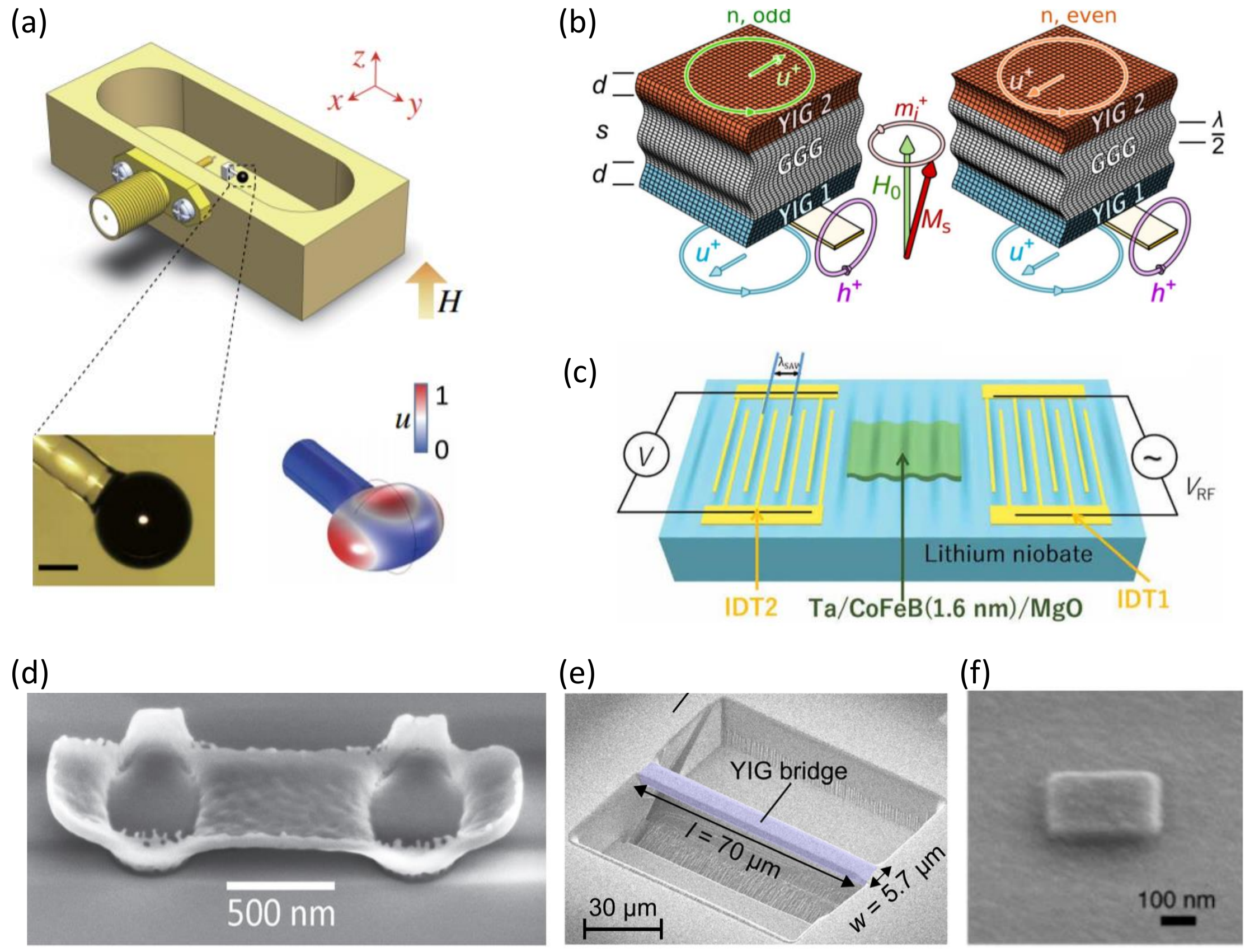}
 \caption{(a) Magnetomechanical device consisting of a YIG sphere inside a 3D microwave cavity. The Kittel magnon mode in YIG is modulated by the mechanical deformation of the sphere. Adopted from Ref.~\onlinecite{2016_SciAdv_Zhang}. (b) Planar YIG device supporting magnon-BAW phonon interaction. BAW phonons in the GGG substrate couple remote magnons modes in the YIG films on the opposite sides of the substrate. Adopted from Ref.~\onlinecite{2020_PRB_An_magnonBAW}. (c) Planar device with a magnetic thin film deposited on top of a piezoelectric substrate (lithium niobate), supporting magnon-SAW phonon interaction. Adopted from Ref.~\onlinecite{2020_SciAdv_Xu_magnonSAW}. (d) Scanning electron microscope (SEM) image of a suspended monocrystalline freestanding YIG bridge fabricated using pulsed laser deposition and lift-off. Adopted from Ref.~\onlinecite{2019_PRAppl_Heyroth}. (e) SEM image of a YIG microbridge structure fabricated using focused ion beam etching. Adopted from Ref.~\onlinecite{2017_APL_Seo}. (f) SEM image of $330 \times 330 \times 30$~nm Ni nanomagnet which supports strong magnon-phonon coupling. Adopted from Ref.~\onlinecite{2019_NC_Berk}.}
 \label{fig_magnemechanics}
\end{figure*}

As of today, most magnetomechanical interactions are reported in macroscopic devices. Recently there have been increasing efforts in pushing magnetomechanical devices towards smaller scales for better device performances or integration.
In spite of the challenges in YIG fabrication, single-crystal free-standing YIG microbeams have been fabricated based on unconventional fabrication approaches such as patterned growth \cite{2019_PRAppl_Heyroth} and angled etching using focused ion beam \cite{2017_APL_Seo,2019_NC_Harii}. As the mass and footprint of the resulting magnetomechanical devices are drastically reduced, one may potentially further enhance the magnon-phonon interaction. By adopting other easy-to-fabricate magnetic materials, magnetomechanical strong coupling has been reported in nanoscale devices \cite{2019_NC_Berk,GodejohannPRB2020}. Moreover, it is proposed that in magnetic nanoparticles, the coupling of magnons with the rotational degree of freedom of the particle can significantly affect the magnon properties, leading to novel opportunities for coherent signal processing \cite{2017_PRB_Keshtgar}.

Developing novel fabrication techniques will dramatically facilitate novel device designs and break many existing experimental restrictions. Therefore, it represents one important future directions for the study on magnon-phonon interactions. Another promising direction involves exploring different materials, such as multiferroics \cite{2007_PRB_Aguilar} and antiferromagnets \cite{2019_JAP_Rezende}, as new magnomechanical platforms. In the context of coherent information processing, it is critical that these materials possess low losses for both magnons and phonons to ensure coherent operations can be performed. Currently most demonstrations of magnetomechanical interactions are still in the classical regime at room temperatures. One straightforward future direction is to bring the magnon-phonon interactions into the quantum regime at cryogenic temperatures, to observe quantum magnetomechanical interactions and perform quantum operations such as entanglement, ground state cooling, squeezing, etc.


Another distinct advantage of microwave magnon modes is their short wavelength and low propagation speed of the dynamic excitations at microwave frequencies compared to electromagnetic waves.  This makes magnons highly beneficial for an efficient miniaturization of microwave components.  Phonon based microwave devices that incorporate piezoelectric materials have similar advantages for miniaturization.\cite{WeigelIEEETMTT2002}  Traditionally, quartz ($\alpha$-SiO$_2$) has been the material of choice for piezoelectric transducers, but more recently materials like LiNbO$_3$ or LiTaO$_3$ are often used for high quality peizoelectric transducres, since they can reach mechanical quality factors approaching $10^5$.\cite{ShaoPRAppl2019}  In addition, SiC and AlN have garnered high interest for piezoelectric devices, since these materials can be readily integrated with thin film growth processes.  Using any of these materials, phonons, such as surface acoustic waves, can be coupled to electromagnetic radiation via interdigitated transducers, such as the ones shown in Figs.~\ref{fig_magnemechanics}(c).  Such surface acoustic wave devices have formed the basis for many contemporary investigations of magnon-phonon coupling.

In fact, more than a century ago the deep connection between spin and mechanical angular momentum has already been established by a pair of pioneering measurements.  Namely, Barnett showed that the mechanical rotation can result in a net magnetization\cite{BarnettPR1915}, while concomitantly Einstein and de Haas demonstrated the inverse phenomenon of mechanical rotation due to magnetization reversal.\cite{EinsteinDPG1915}  This provides in principle a pathway for coupling phonons to magnetization and its dynamics\cite{ZhangPRL2014}.  Recently, experiments with elastically driven ferromagnetic resonance in Ni$_{81}$Fe$_{19}$ on LiNbO$_3$ suggest that such direct angular momentum coupling is indeed possible.\cite{KobayashiPRL2017}  However, a more efficient pathway for magnon-phonon interactions is due to magneoelastic coupling\cite{TurovFMM1956} and it was already pointed out early on that the classical understanding of this coupling provides the same dynamic equations as a quantum mechanical description.\cite{KittelPR1958}  This magnetoelastic coupling gives rise to hybrid magnetoelastic modes or fully mixed magnon-polarons, when the energy of the phonons coincides with the corresponding magnon enmergies.  Since both phonons and magnons also couple directly to optical photons, details of their interactions can therefore be directly investigated via spatially resolved inelastic light scattering.\cite{ZhaoArxiv2020}  Furthermore, magnon-polarons may also be important for understanding spin currents interacting with temperature gradients,\cite{BozhkoLTP2020} {\em e.g.}, for spin Seebeck effects \cite{FlebusPRB2017,CornelissenPRB2017}.

More importantly, magnetoelastic coupling provides new opportunities for coherent interactions between magnons and phonons.  An important starting point was the demonstration that ferromagnetic resonance in Ni can be directly excited via surface acoustic waves in LiNbO$_3$.\cite{WeilerPRL2011}. Theoretically, it was shown that the coherent elastic interfacial excitation of the magnetization dynamics may be associated with the emergence of evanescent interface states, which depend on the relative orientation of the magnetization with respect to the phonon propagation.\cite{KameraPRB2015}  In addition it was shown that the phonon propagation can be modulated due to the magnetoelastic interactions in complex ways.  Namely, the coupling to the magnons may give rise to a mixing between different phonon modes, but the precise interactions depend very sensitively on processing conditions that change the interfacial microstructure.\cite{ZhaoPRAppl2020}

As the complex modulation of phonon propagation in the presence of magnetoelastic interaction shows, it is also possible to generate phonons from magnons.  This was directly demonstrated by combining excitations of ferromagnetic resonance in Ni wire through {\em rf} magnetic fields with detection through surface acoustic waves in adjacent LiNbO$_3$ via interdigitated transducers.\cite{BhuktareSR2017}  Interestingly, using inealstic light scattering it was possible to directly measure the angular momentum associated with phonons generated from magnons.\cite{HolandaNP2018}  Using an adiabatic conversion of magnons into phonons via spatially varying magnetic fields\cite{SchloemannJAP1964} enabled to separate the magnon and phonon components of the hybridized modes.  Furthermore, the higher group velocity of phonons can be detected in pulsed time-dependent measurements and provides thereby direct verification of the magnon to phonon conversion.  Similarly, hybrid magnon-phonon modes may therefore provide significantly faster magnon propagation.\cite{SatohNP2012,ShenPRL2015}  Thus phonons provide new pathways for fast angular momentum transport.\cite{JungfleischNP2018}  Similarly, due to their long coherence length, phonons may also mediate magnon coupling over long distances.  This was demonstrated by coupled magnon modes for two 200-nm thick yttrium iron garnet films deposited on opposite sides of a 0.5-mm thick gadolinium gallium garnet substrate.\cite{AnPRB2020}  Gadolinium gallium garnet has a very long phonon mean free path, such that the phonon decay length is about 2 mm and therefore exceeds the thickness of the substrate.  This results in standing phonon modes that give rise to constructive or destructive interference between the ferromagnetic resonance of the two yttrium iron garnet layers depending on whether the phonon mode number is odd or even.  This shows that phonons can provide a coherent long distance pathway for angular momentum communication, which may have direct relevance for quantum transduction.

Another important aspect for quantum devices is non-reciprocity, which allows phase coherent information processing with a well defined directionality.  In magnetic systems, non-recirpocal phenomena are a direct consequence of the inherently broken time-reversal symmetry, and has been harnessed for many microwave and optical devices, such as circulators and isolators.  However, if these systems are based on coupling electromagnetic waves to the magnetic system they become invariably bulky.  Here the coupling of magnons to phonons provides a distinct opportunity for combining a compact device structure with high efficiency performance.  Already in the 1970's surface acoustic wave insulators were demonstrated by using yttrium iron garnet films on gadolinium gallium garnet.\cite{LewisAPL1972}  More recently, similar non-reciprocal propagation of surface acoustic waves has also been observed for LiNbO$_3$ combined with Ni.\cite{SasakiPRB2017}  It turns out that interfacial chiral exchange coupling, the Dzyaloshinskii-Moriya interaction, can shift the hybridization gaps relative to each other for the magnon-phonon modes with opposite momenta and thus allows to optimize the non-reciprocity.\cite{VerbaPRAppl2018}  Further improvements are possible by increasing the magnetoelastic coupling, and towards this end an isolation of close to 50~dB near 1.5~GHz was recently demonstrated by combining LiNbO$_3$ with FeGaB.\cite{ShahArxiv2020} 
\begin{figure}[htb]
 \centering
 \includegraphics[width=3.0 in]{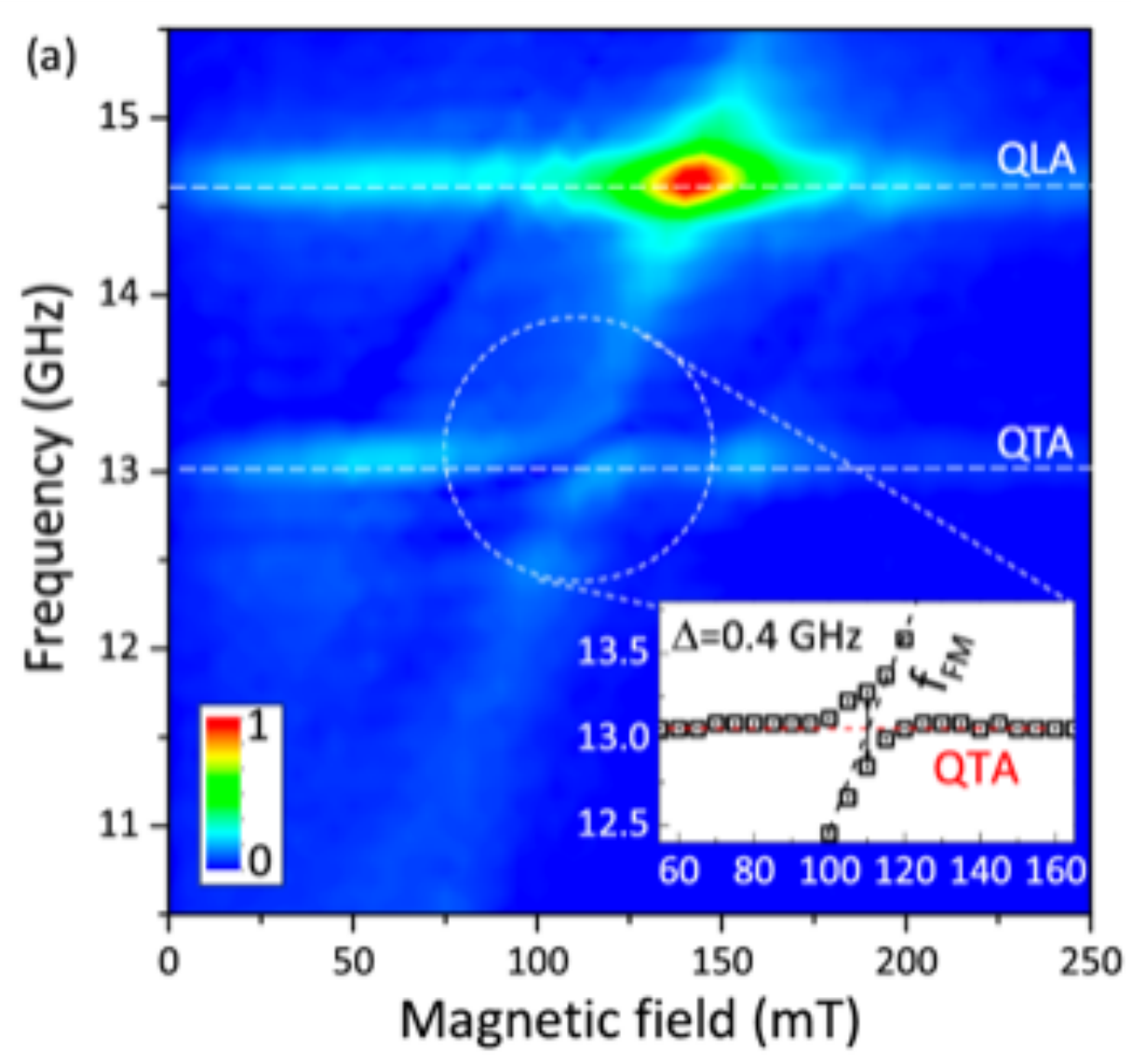}
 \caption{Hybridization of magnon and phonon modes in the strong coupling regime.  The color map shows the spectral density of transient Kerr rotation signals as a function of the external magnetic field.  The dispersing mode is the ferromagnetic resonance of the Fe$_{81}$Ga$_{19}$ crossing the quasi transverse (QTA) and quasi longitudinal acoustic (QLA) modes of the nanograting.  The inset shows the clear hybridization gap of the ferromagnetic resonance with the QTA mode.  Adapted from Ref.~\onlinecite{GodejohannPRB2020}.}
 \label{fig_strong-magnon-phonon}
\end{figure}

The fact that strongly magnetostrictive materials provide superior performance for non-reciprocal devices, exemplifies the opportunities that still remain in optimizing magnon-phonon coupled devices.  The majority of investigations have focused on Ni based devices.  Ni has moderately high magneto-elastic coupling, but also very high magnetic damping.  This means that it is very difficult to fabricate Ni-based systems in the strong magnon-phonon coupling regime.  Nevertheless, even with Ni it is possible to have magnon-phonon coupled systems where the cooperativity approaches and start to exceed 1.\cite{BerkNC2019}  However, it turns out that Fe$_{81}$Ga$_{19}$ can outperform Ni significantly, since it combines strong magnetoelastic coupling with reasonably low magnetic damping.  Using periodic thickness modulation for increasing the magnetoelastic coupling to specific phonon modes, Godejohann {\em et al.}\ showed that clear magnon polarons in the strong coupling regime can be observed,\cite{GodejohannPRB2020} see Fig.~\ref{fig_strong-magnon-phonon}.  These measurements showed relatively long coherence times with decoherence rates of 30~MHz for the phonons and 170~MHz for the magnons.  Together with the experimentally measured coupling of 200~MHz this results in a cooperativity exceeding 7.8.  This shows that a judicious choice of magnetic materials offers still plenty of opportunities to improve device performance.  In that respect, towards the goal of getting stronger magnetoelastic coupling with a simultaneously reduced magnetic damping, amorphous B- and C-doping of FeGa and FeCo alloys appears to be a promising direction.\cite{LiangSensor2020}


\subsection{A Brief Theoretical Consideration: Magnon-Mediated Operations via Dynamic Tuning}

\begin{figure*}%
\includegraphics[width=161mm]{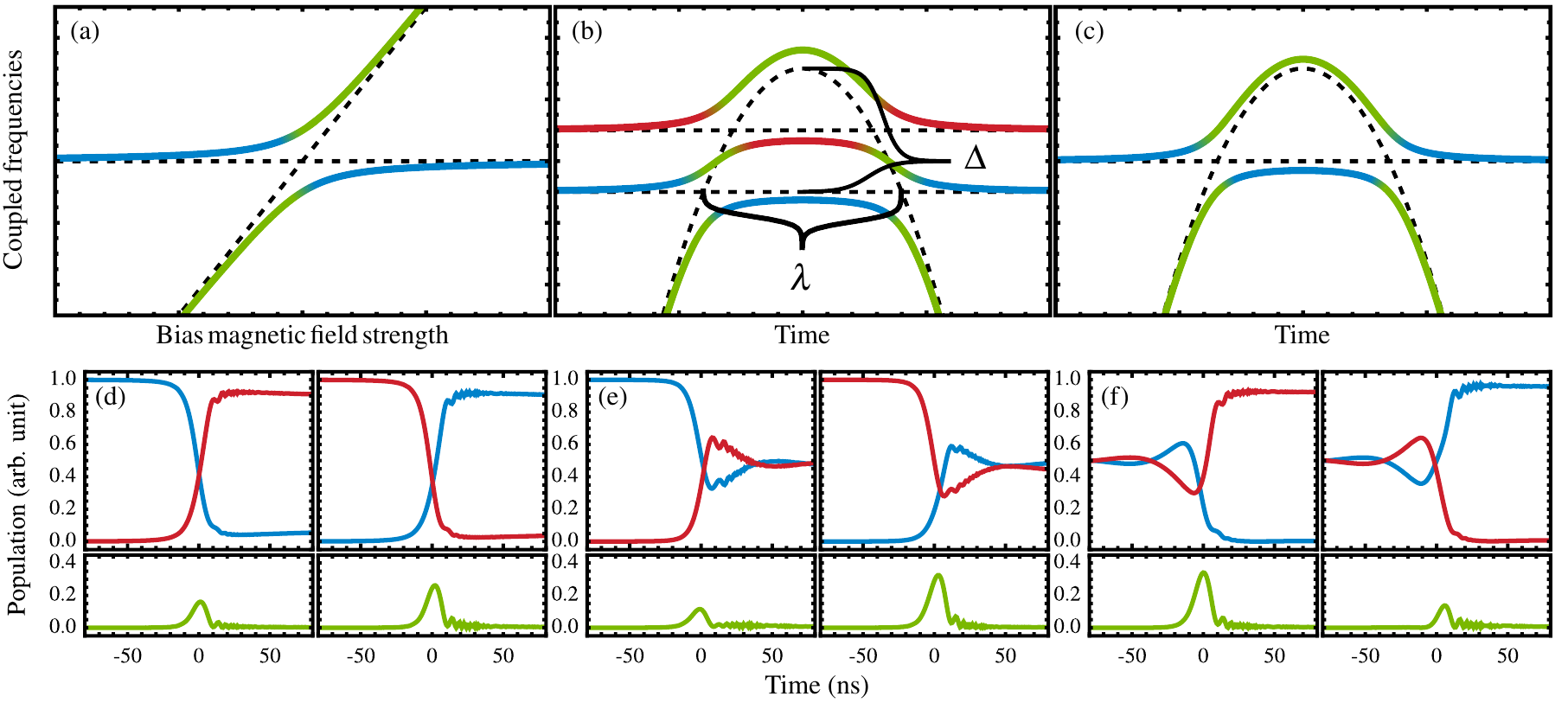}%
\caption{\label{fig:mfigTT}(a) Dependence of the coupled mode frequencies $\omega_{\pm}$ Eqs.~\eqref{eq:1to1soln} on the bias magnetic field strength (i.e., magnonic mode frequency $\omega_\mathrm{m}$). Dashed lines show the frequencies of uncoupled photonic $\omega_{1}$ and magnonic $\omega_{\mathrm{m}}$ modes. The color of the branches denotes the distribution of the coupled mode energy between the magnonic (green) and photonic (blue) resonators. (b) Time dependence of the instantaneous eigen-frequencies of Eqs.~\eqref{eq:2to1} for a parabolic profile Eq.~\eqref{eq:para} of the bias magnetic field. The parameter $\Delta$ determines the maximum value of the magnon frequency $\omega_{\mathrm{m}}$, while $\lambda$ is proportional to the curvature of the parabola. The color of the branches denotes the distribution of the coupled mode energy between the interacting resonators. (c) Time dependence of the instantaneous eigen-frequencies of Eqs.~\eqref{eq:1to1} for a parabolic profile Eq.~\eqref{eq:para} of the bias magnetic field. The color of the branch denotes the content of the mode (green – magnonic mode $a_\mathrm{m}$, blue – photonic $a_{1}$ mode). (d) Temporal profiles of the asymmetric SWAP operation using a parabolic pulsed magnetic field profile with $\lambda=2.0\,\kappa_\mathrm{c}^3$, $\Delta=-1.0\,\kappa_\mathrm{c}$, and $\Delta\omega_{1,2}=\omega_{2}-\omega_{1}=2\pi\cdot2.0$~MHz. Left panels: information originally contained in $a_{1}$ mode (blue) is coherently transduced to $a_{2}$ mode (red). Right panels: information originally contained in $a_{2}$ mode (red) is coherently transduced to $a_{1}$ mode (blue). Bottom panels show the population of the magnonic mode $a_\mathrm{m}$ during the process. (e) Temporal profiles of the populations of the interacting modes during the \textcolor{black}{SPLIT} operation realized with a parabolic field profile using $\lambda=1.625\,\kappa_\mathrm{c}^3$, $\Delta=-0.985\,\kappa_\mathrm{c}$, and $\Delta\omega_{1,2}=2\pi\cdot12$~MHz. Left panels: information originally contained in $a_{1}$ (blue) mode is split into $a_{1}$ and $a_{2}$ (red) modes. Right panels: information originally contained in $a_{2}$ mode is split between $a_{1}$ and $a_{2}$ modes. Bottom panels show the population of the magnonic mode $a_\mathrm{m}$. (f) Temporal profiles showing the reversibility of the \textcolor{black}{SPLIT} operation while using the same parameters of the field pulse as in Fig.~\ref{fig:mfigTT}(e). Initial states in the simulations correspond to the final states of the simulations shown in Fig.~\ref{fig:mfigTT}(e). In both cases (left and right panels) the initial state is an entangled state with different phase relations between partial $a_{1}$ and $a_{2}$ photonic modes for two panels. Repeating the \textcolor{black}{SPLIT} operation successfully disentangles the states into a pure $a_{2}$ (left panel) or $a_{1}$ (right panel) state.}%
\end{figure*}

Hybrid magnonic systems benefit from the wide-ranged tunability of the resonant magnon frequency using changes in bias magnetic field strength. Though existing systems achieving high coupling rates encompass a wide variety of geometries ranging from millimeter-scale microwave cavities \cite{ZhangPRL2014,TabuchiPRL2014,TabuchiScience2015} to micrometer-scale superconducting coplanar resonators \cite{HueblPRL2013,LiPRL2019,HouPRL2019}, all of these systems function via resonant coupling between the electromagnetic and magnonic resonators. This resonant coupling occurs when the resonant magnon frequency is near the resonant frequency of the photonic resonator. Changing the strength of the bias magnetic field and, thus, the magnon resonant frequency allows one to continuously tune the hybrid system on and off the resonance and observe the famous anti-crossing behavior of the coupled oscillator’s frequencies, shown schematically in Fig.~\ref{fig:mfigTT}(a).

The behavior of a hybrid magnon-photon system can be mathematically described by a simple system of equations for complex amplitudes of the photonic $a_{1}(t)$ and magnonic $a_\mathrm{m}(t)$ resonator modes \cite{intermag2020,mmm2020}:

\begin{equation}
\begin{aligned}
\frac{da_{1}}{dt}+i\omega_{1}a_{1}&=-i\kappa_{1}a_\mathrm{m},\\
\frac{da_{m}}{dt}+i\omega_\mathrm{m}a_\mathrm{m}-\Gamma_\mathrm{m}a_\mathrm{m}&=-i\kappa^{*}_{1}a_{1}.
\label{eq:1to1}
\end{aligned}
\end{equation}

\noindent Here $\omega_{1}$ and $\omega_\mathrm{m}$ are the resonant frequencies of the photonic and magnonic modes, respectively; the magnonic frequency $\omega_\mathrm{m}$ depends on the applied bias magnetic field and can be easily tuned in a wide range. $\Gamma_\mathrm{m}$ is the damping rate of the magnonic resonator; the corresponding term was, for simplicity, neglected in the equation for photonic mode $a_{1}$ since the photon damping rate $\Gamma_\mathrm{p}$ is usually much lower than $\Gamma_\mathrm{m}$. The parameter $\kappa_{1}$ in Eqs.~\eqref{eq:1to1} is the coupling rate between the magnon and photon modes and can exceed 100 MHz \cite{HueblPRL2013,ZhangPRL2014,TabuchiPRL2014,TabuchiScience2015,LiPRL2019, HouPRL2019}.

Model Eqs.~\eqref{eq:1to1} describe linear coupling of two resonant modes. One can easily find eigen-frequencies of two coupled oscillations formed as a result of mode interaction:

\begin{equation}
\omega_\pm=\frac{\omega_{1}+\omega_\mathrm{m}}{2}\pm\sqrt{\left(\frac{\omega_{1}-\omega_\mathrm{m}}{2}\right)^{2}+{\kappa_{1}}^{2}}.
\label{eq:1to1soln}
\end{equation}
\noindent The solutions for Eq. \eqref{eq:1to1soln} for varying $\omega_\mathrm{m}$ (which, physically, corresponds to varying the bias magnetic field) are shown in Fig.~\ref{fig:mfigTT}(a) and demonstrates the formation of a mode anti-crossing at the point of the strongest resonant coupling ($\omega_{1}=\omega_\mathrm{m}$), where the modes become strongly hybridized to prevent degeneracy of their frequencies. The gap between the coupled mode frequencies at the anti-crossing point is proportional to the coupling strength, $\omega_{+}-\omega_{-}=2\lvert{\kappa_{1}\rvert}$.

All the existing studies of hybrid magnon-photonic systems focus on quasi-static tuning of the magnon resonance frequency $\omega_\mathrm{m}$, in which the characteristic time $\tau$ of magnetic field variation is much longer than the magnon lifetime, $\tau\gg1/\Gamma_\mathrm{m}$. In this case, the hybrid system behaves as an usual resonance system with statically tunable resonance parameters described by Eq.~\eqref{eq:1to1soln}.

The strongly coupled ($\lvert{\kappa_{1}\rvert}\gg\Gamma_\mathrm{m}$) magnon-photon systems, however, may exhibit much richer and much more interesting behavior in case of dynamic tuning of the resonance frequency $\omega_\mathrm{m}$. If the characteristic time of the magnetic field variation $\tau$ satisfies $1/\Gamma_\mathrm{m}\gg\tau\gtrsim1/\lvert{\kappa_{1}\rvert}$, the parameters of the magnon mode change during the lifetime of a single magnon; at the same time, magnon and photon modes interact over a sufficiently long time interval ($\tau\lvert{\kappa_{1}\rvert}\gtrsim1$) to ensure efficient energy and information exchange between the modes.

In the simplest case of a linearly ramped bias magnetic field $B(t)=B_{0}+(dB/dt)t$ the behavior of the dynamically-tuned hybrid system can be qualitatively understood from Fig.~\ref{fig:mfigTT}(a), in which the horizontal axis now plays the role of time. As the bias field $B(t)$ comes closer to the resonance value, the two modes hybridize, but the excitation initially localized within the higher (lower) mode, will continue to stay in that mode. The overall result of the “passage” shown in Fig.~\ref{fig:mfigTT}(a) is the transfer of energy and information from the photonic mode to the magnonic one [along the upper branch $\omega_{+}(t)$], and vice versa [along the lower branch $\omega_{-}(t)$]. Note, however, that this simple picture of an \textit{adiabatic passage} is, strictly speaking, quantitatively correct for relatively slow ramps $\tau\gg\lvert{\kappa_{1}\rvert}$; in a more realistic non-adiabatic setting $\tau\sim1/\lvert{\kappa_{1}\rvert}$ a more elaborate model of hybrid system dynamics, which is described below, must be utilized.

Mathematically, dynamically-tuned hybrid systems can be described by the system of equations similar to Eqs.~\eqref{eq:1to1}, but with time-dependent magnon frequency $\omega_\mathrm{m}(t)$, determined by the time profile of the bias magnetic field. By shaping the profile of the pulsed magnetic field one can control how strongly, when, and for how long the resonant coupling occurs, which, in turn, will determine the overall outcome of the “passage”.

Below, we illustrate some of the capabilities of dynamically-tuned hybrid magnon-photon systems using, as an example, a system that consists of two photonic resonators (amplitudes $a_{1}$ and $a_{2}$, resonant frequencies $\omega_{1}$ and $\omega_{2}$) and one magnonic resonator (amplitude $a_\mathrm{m}$, frequency $\omega_\mathrm{m}$, damping rate $\Gamma_\mathrm{m}$). The photonic resonators are not coupled directly, but both interact with the magnonic resonator with coupling rates $\kappa_{1}$ and $\kappa_{2}$, respectively. The equations describing this system is a direct generalization of Eqs.~\eqref{eq:1to1}:

\begin{equation}
\begin{aligned}
\frac{da_{1}}{dt}+i\omega_{1}a_{1}&=-i\kappa_{1}a_\mathrm{m},\\
\frac{da_{m}}{dt}+i\omega_\mathrm{m}(t)a_\mathrm{m}-\Gamma_\mathrm{m}a_\mathrm{m}&=-i\kappa^{*}_{1}a_{1}-i\kappa^{*}_{2}a_{2},\\
\frac{da_{2}}{dt}+i\omega_{2}a_{2}&=-i\kappa_{2}a_\mathrm{m}.
\label{eq:2to1}
\end{aligned}
\end{equation}

In the following simulations we used conservative estimate for the coupling rates, $\kappa_{1,2}=\kappa_\mathrm{c}=2\pi\cdot20$~MHz, which is much lower than the experimentally achievable values. We assumed the damping value $\Gamma_\mathrm{m}=6.28$~$\mu$s$^{-1}$ typical for magnonic resonators made of yttrium iron garnet.

The dynamics of the hybrid magnon-photon system is determined by the profile of the pulsed bias magnetic field, i.e., by the time dependence of the magnonic resonance frequency $\omega_\mathrm{m}(t)$. Here we consider a parabolically shaped profile,

\begin{equation}
\omega_\mathrm{m}(t)-\omega_{1}=-\lambda t^{2}+\Delta,
\label{eq:para}
\end{equation}
\noindent where the parameters $\Delta$ and $\lambda$ determine the maximum value and the curvature $\lambda$ of the pulse as shown in Fig.~\ref{fig:mfigTT}(b). The characteristic passage time in this case is $\tau\sim\sqrt{\kappa_\mathrm{c}/\lambda}$.

In the adiabatic limit $\tau\gg\lvert{\kappa_\mathrm{c}}\rvert$ the parabolic “passage” has no effect on the final populations of the interacting modes, as the magnonic frequency $\omega_\mathrm{m}(t)$ crosses each of the photonic modes twice. The situation, however, is different in the non-adiabatic regime $\tau\sim1/\kappa_\mathrm{c}$ , in which the final populations of each mode depend on the pulse parameters $\lambda$ and $\Delta$. This means that by simply changing the shape of the pulsed magnetic field profile, different operations can be realized in the same physical system.

One such operation useful in quantum computing is the coherent exchange of information between the two photonic modes (SWAP operation). Fig.~\ref{fig:mfigTT}(d) shows a simulation of this SWAP operation realized using $\lambda=2.0\,\kappa_\mathrm{c}^{3}=2\pi\cdot0.63$~MHz/ns$^{2}$, $\Delta=-1.0\,\kappa_\mathrm{c}=-2\pi\cdot20.0$~MHz, and slightly different frequencies of photonic resonators $\Delta\omega_{1,2}=\omega_{2}-\omega_{1}=2\pi\cdot2.0$~MHz. Interestingly, the SWAP operation is asymmetric for a parabolic ramp, due to the non-zero resonant frequency separation $\Delta\omega_{1,2}$ between the photonic modes. This asymmetry in the SWAP operation can clearly be seen in the content of the magnonic mode during passage, shown in the bottom panels of Fig.~\ref{fig:mfigTT}(d). It is important to note, that this SWAP operation is also possible in the case of degenerate photonic modes.

It is also important to note, that the population of the magnon mode is zero after the passage, i.e., the magnonic serves only as a mediator of energy transfer between the photonic subsystems. Since the passage occurs during a relatively short time $\tau\ll1/\Gamma_\mathrm{m}$, the magnon dissipation does not play a crucial role in this process, and the accuracy of the SWAP operation is close to 100\%.

Another useful operation in quantum computing is the \textcolor{black}{SPLIT} operation, where one coherent information state is split into a coherent superposition of multiple information states. Such \textcolor{black}{SPLIT} operation can be performed using the same parabolic-shaped pulsed magnetic field profile Eq.~\eqref{eq:1to1soln} with different values of the curvature $\lambda$ and maximum frequency $\Delta$. Simulation of this operation in Fig.~\ref{fig:mfigTT}(e) used $\lambda=1.625\,\kappa_\mathrm{c}^{3}=2\pi\cdot0.513$~MHz/ns$^{2}$, $\Delta=-0.985\,\kappa_\mathrm{c}=-2\pi\cdot19.7$~MHz, and $\Delta\omega_{1,2}=\omega_{2}-\omega_{1}=2\pi\cdot12.0$~MHz shows an asymmetry similar to that seen in the SWAP operation especially pronounced in the time dependence of the magnonic mode. Interestingly, despite this observable asymmetry the \textcolor{black}{SPLIT} operation may be reversed by simply repeating the same operation on the coherent superposition of information states, shown in Fig.~\ref{fig:mfigTT}(f). This \textcolor{black}{SPLIT} operation and its reversal can also be performed on degenerate photonic modes.

Dynamic tuning of resonant magnon frequencies may also be useful for investigation of fundamental properties of single magnons. For example, one can consider a system of a single photonic resonator coupled to a magnonic resonator as described in the mathematical model Eqs.~\eqref{eq:1to1} with a parabolic profile $\omega_\mathrm{m}(t)$ from Eq.~\eqref{eq:para}. The dependence of the instantaneous eigen-frequencies of the hybrid modes for this case is shown in Fig.~\ref{fig:mfigTT}(c). If the magnetic field profile is non-adiabatic, $\tau\sim\lvert{\kappa_\mathrm{c}}\rvert$, then only a partial (say, 50\%) exchange of quantum state will occur at each of the two points of resonant hybridization. Then, in the time interval between the hybridization points the system will exist in an entangled magnon/photon state. The two components of this state will interfere at the second hybridization point, and the final population of each mode will depend on details of single-magnon dynamics. Experiments of this type will be useful to investigate single-magnon non-dissipative decoherence processes, which are difficult to study using other means.

Finally, we note, that similar magnon-mediated coherent operations, described here for the case of photonic modes, are also possible with phonons and other solid-state excitations that efficiently couple to magnons. Thus, magnons represent a promising candidate for a universal mediator of coherent quantum information transduction in heterogeneous quantum systems.

\section{Directions in Using Layered ``Synthetic Magnets" for Quantum Magnonics}

Magnon-photon and magnon-phonon devices inherently feature a two-dimensional circuit layout due to the physical nature of their hybridizing mechanisms. Another emerging direction in the context of quantum magnonics takes advantage of the magnon-magnon coupling mechanism induced by exchange-coupled, synthetic magnetic layers. 

\begin{figure*}[ht]
 \centering \includegraphics[scale = .58]{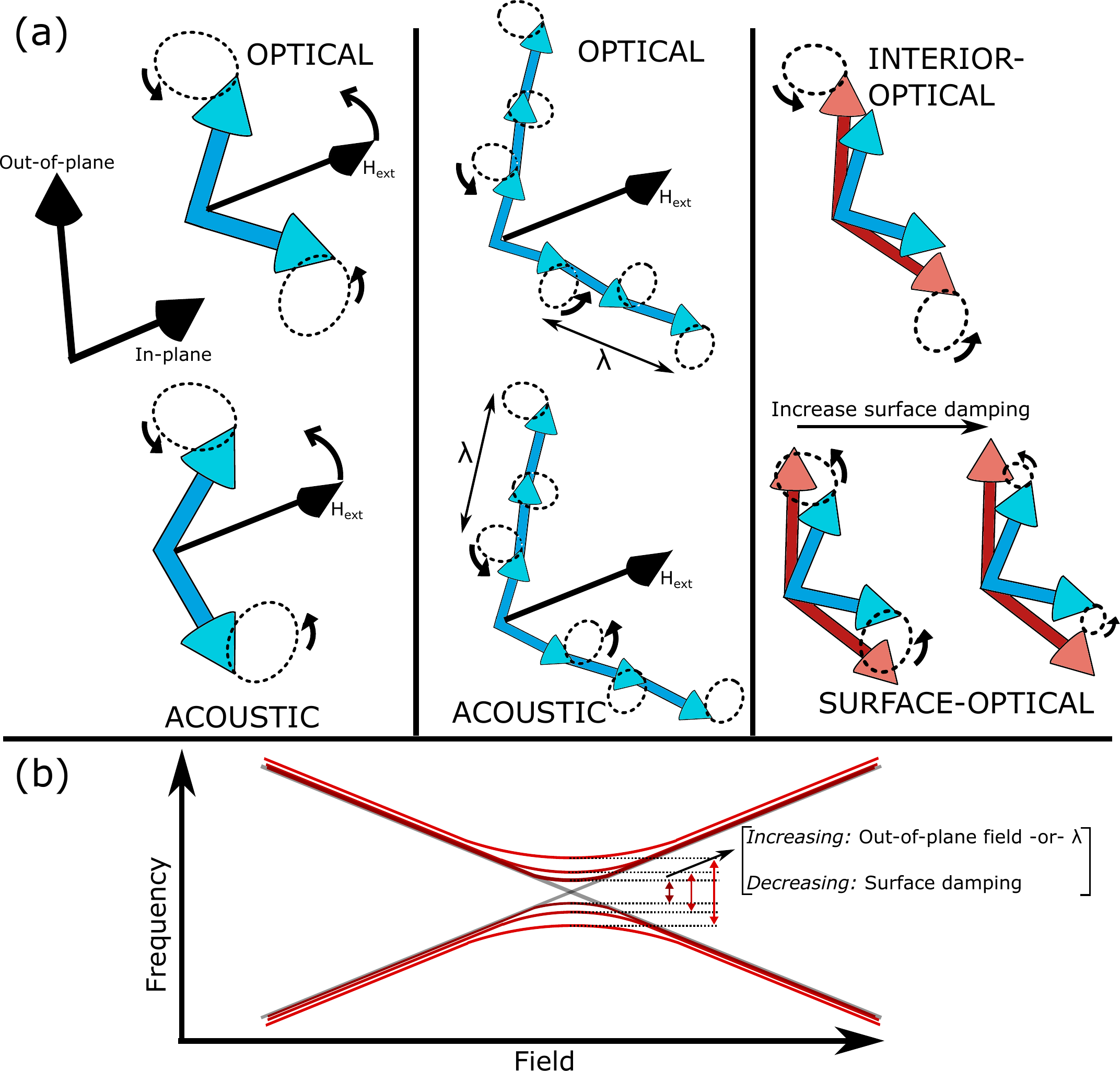}
 \caption{ (a) Illustration of three pairs of magnonic excitations that can be coupled together within a layered antiferromagnet.  In the context of synthetic antiferromagnets, blue arrows represent magnetization directions two magnetic layers.  Alternatively, in a bulk sample, each arrow represents the magnetization of two magnetic sublattices.   The red arrows are used in the case of a layered magnetic material with \textit{four} magnetic layers, and represent the interior layers while the blue arrows represent the surface layers.  From left-to-right the following pairs of magnons can be coupled together:  1) A spatially uniform optical and acoustic magnon, 2) A finite wavenumber optical and acoustic magnon, 3) An optical magnon residing on the interior and surfaces of a layered system comprised of four magnetic layers.  (b) In these three instances, the magnon-magnon interaction can be adjusted by rotating an external field out-of-plane, changing the wavenumber, or changing the magnetic damping on the surface layers relative to the interior layers.   \label{O_A_coupling}}
\end{figure*}

Recently, strong light-matter interaction became a key basis for many experiments in areas of quantum information technologies, including processing, storage or sensing \cite{LachanceQuirionAPEx2019, Harder2018, Xiang2013, Wallquist2009, Grezes2016, Bienfait2016, Eichler2017,xiongyz_nsep2020}. Magnetically-ordered materials are promising candidates for reaching the strong coupling regime of coherent excitation exchange, due to their high spin densities. Antiferromagnetic systems intrinsically possess two magnon modes, typically referred to as acoustic and optical modes \cite{Keffer1952, Rezende2019} due to their sub-lattice natures. Their dynamics is characterised by the strength of the exchange coupling and antiferromagnetic crystals tend to show a large coupling strength, bringing the frequency of their dynamics into THz regions. In this regard, synthetic antiferromagnets\cite{Grunberg1986, Duine2018} and layered antiferromagnets\cite{Mak2019} are especially appealing in terms of accessibility since their weaker interlayer exchange coupling nature allows GHz resonances\cite{Krebs1990, Zhang1994, Heinrich2003, Seki2009, Timopheev2014, Tanaka2014, Liu2014} which can also be easily tuned by growth/material parameters \cite{Liu2014, Sorokin2020}.   As illustrated in Fig.~\ref{O_A_coupling}, this Section will broadly be concerned with providing an overview of how the magnon-magnon coupling mechanism can be controlled layered antiferromagnetic materials.    

\subsection{\textcolor{black}{Magnon-Magnon Coupling in Synthetic Antiferromagnets}}

\begin{figure*}[ht]
 \centering \includegraphics[]{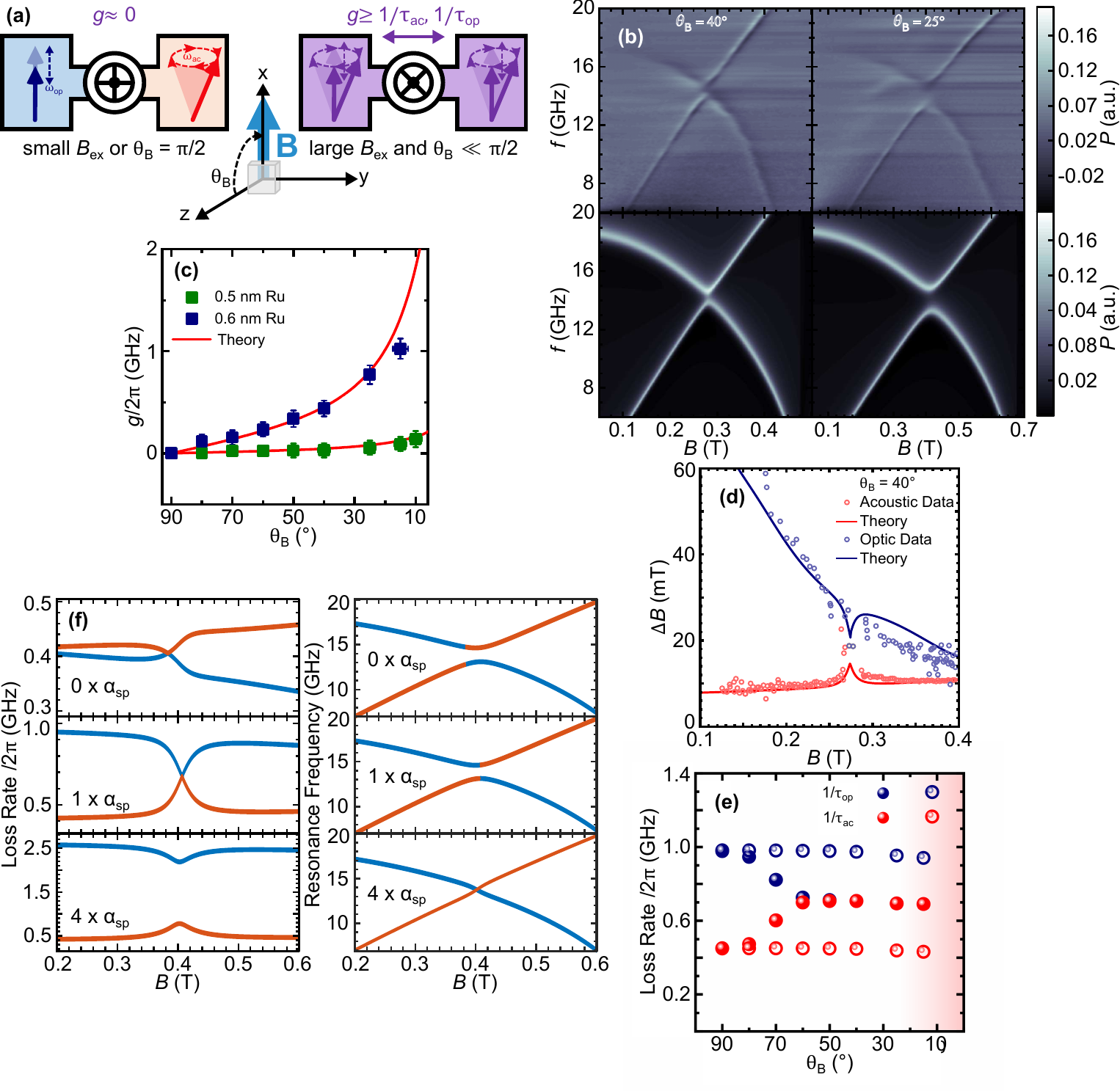}
 \caption{(a) Schematic of the coupling process between acoustic and optic magnon modes. The coupling strength is controlled by engineering the exchange coupling between the magnetic layers and by dc magnetic field orientation. The coordinate systems shows the definition of the magnetic field angle $\theta_{\mathrm{B}}$. (b) Measured and simulated microwave transmission through a CPW with a SAF on top as a function of microwave frequency and dc magnetic field. The two columns are presenting the results for different magnetic field angles. (c) Extracted coupling strength $g$ as a function of the dc magnetic field angle $\theta_{\mathrm{B}}$ for two different spacer layer thickness between the ferromagnetic layers of the SAF. (d) Magnetic field linewidth $\Delta B$ of the acoustic and optic magnon modes as a function of dc magnetic field, applied at an angle of $4^\circ$. At the point of degeneracy at $B \approx 275~\mathrm{mT}$ the linewidth are merging towards a single value. (e) Calculated loss rates of the acoustic and optic magnon modes as a function of the dc magnetic field angle $\theta_{\mathrm{B}}$. The solid symbols are results from calculations assuming a coupling, while the empty symbols show results without a coupling. (f) Simulated resonance frequencies and loss rates of the acoustic and optic modes as a function of the dc magnetic field for $\theta_{\mathrm{B}} = 27^\circ$. Parameters are the same as presented in Ref.~\onlinecite{Sud2020}, where only the damping contribution from mutual spin pumping $\alpha_{\mathrm{sp}}$ is changed. The simulation results suggest that mutual spin pumping is partially contributing to the coupling process. \label{fig:1}}
\end{figure*}

We here review recent progress of magnon-magnon coupling of uniform spin-wave modes in synthetic antiferromagnets\cite{Sud2020}. To hybridize the acoustic and optical modes, the twofold rotational symmetry of the two magnetisation vectors about the dc magnetic field needs to be broken\cite{MacNeill2019}. This is , \textcolor{black}{e.g.}, by applying a dc magnetic field \textcolor{black}{($B_{\mathrm{0}}$)} at an angle $\theta_{\mathrm{B}}$ with respect to the \textcolor{black}{film plane;} see the schematic coordinate system in Fig.~\ref{fig:1}(a). The demagnetization field \textcolor{black}{($B_{\mathrm{s}}$)} arising from the thin-film nature prevents a full alignment of the magnetisation to the applied dc magnetic field. This breaks the rotational symmetry and enables an interaction between the two magnon modes \cite{MacNeill2019,Sud2020}. Furthermore, the strength of the interaction increases as the out-of-plane angle increases. This introduces an in-situ control over the coupling strength between the two antiferromagnetic modes \cite{MacNeill2019, Shiota2020, Sud2020}. 

In the case of a small interlayer exchange field $B_{\mathrm{ex}}$ or for the in-plane condition, the acoustic and optic modes [see Fig.~\ref{fig:1}(a)] cannot interact strongly at their point degenerate a energy, showing a mode crossing. This degeneracy can be lifted by having large $B_{\mathrm{ex}}$ and a small $\theta_{\mathrm{B}}$, where an avoided crossing starts to emerge as a signature of mode hybridisation, as shown in Fig. ~\ref{fig:1}(b). In this regime, the energy transfer between the two modes takes place at the rate of $g$/2$\pi$ within the magnon system. Here, the engineering of $B_{\mathrm{ex}}$ and in-situ $\theta_{\mathrm{B}}$ tuning acts as a valve of the mode coupling strength as depicted in Fig.~\ref{fig:1}(a). 

Sud et al. experimentally show strong magnon-magnon coupling in a CoFeB/Ru/CoFeB synthetic antiferromagnets\cite{Sud2020}. Figure \ref{fig:1}(b) represents spin-wave spectra measured by using a broad-band microwave transmission line with the sample placed on top of it, as a function of microwave frequency and dc magnetic field. For smaller $\theta_{\mathrm{B}}$ a clear gap is showing at the degeneracy point of the acoustic and optic modes. The extracted coupling strength $g$ as a function of the magnetic field angle is shown in Fig.~\ref{fig:1}(c) for two different thickness of the non-magnetic Ru layer, which shows excellent agreement with the following equation derived from the the LLG equation for coupled magnetic moments at the macrospin limit\cite{Sud2020}:
\begin{equation}\label{eq:couplingSyAF}
g\mathrm{=}\frac{\gamma B_{\text{ex}}B_0}{2B_\text{s}+4B_{\text{ex}}}{\mathrm{cos} {\theta }_\text{B}}. 
\end{equation}
As shown in this equation, $g$ is maximised for small $\theta_{\mathrm{B}}$ and large $B_{\text{ex}}$, which are well-demonstrated by their experiments.

As the resonance frequencies of the modes experience a level splitting during coupling, the magnetic field linewidth $\Delta B$ shows an attraction. Figure~\ref{fig:1}(d) plots $\Delta B$ as a function of the dc magnetic field for an angle of $40^\circ$. At the point of interaction ($B  \approx 275~\mathrm{mT}$) the linewidth are moving together. Similarly, the frequency linewidth or loss rates $({1}/{\tau_{\mathrm{ac}}})$ and $({1}/{\tau_{\mathrm{op}}})$ are attracted and eventually average at the point of degeneracy, as the coupling becomes stronger [see Fig.~\ref{fig:1}(e)]. This reflects that in the coupled state the acoustic and optic mode no longer are individual and form a single hybrid system with a combined loss rate \cite{Sud2020}. Similar results were also found in magnetic hybrid structures \cite{Li2020}. 

The magnon-magnon coupling can be modelled by considering two coupled Landau-Lifshitz-Gilbert equations, where specifically a mutual spin pumping is assumed \cite{Chiba2015, Sud2020}. The mutual spin pumping describes a spin current exchange between the two ferromagnetic layers and adds an additional damping to the dynamics, apart from the Gilbert damping. This term is crucial to model the experimental findings accurately, especially the loss rates of the hybrid system. Notably, extended simulation work suggests that the magnon-magnon coupling is, partially, mediated by spin currents \cite{Sud2020}, as shown in Fig.~\ref{fig:1}(f). The simulated plots depict resonance frequencies and loss rates as a function of dc magnetic field at an angle of $27^\circ$ for different values of the damping contribution from spin pumping $\alpha_{\mathrm{sp}}$. The results suggest that an increased damping due to spin pumping can lead to a resonance frequency crossing and a loss rate splitting, similar as in cavity magnonics\cite{Wang2020}. However, further experimental studies are required to fully confirm this.

In conclusion, synthetic antiferromagnets are highly promising for novel spin-wave states in spintronic devices \cite{Chumak2015}, high speed information processing and storage \cite{Jungwirth2016, Baltz2018, Duine2018, Olejnik2018}, due to their precisely engineerable properties and in-situ tuning capabilities. Together, these material systems are adaptive for future applications.

\subsection{\textcolor{black}{Magnon-Magnon Coupling in 2D van der Waals Materials}}

\begin{figure*}[ht]
 \centering \includegraphics[width=18 cm]{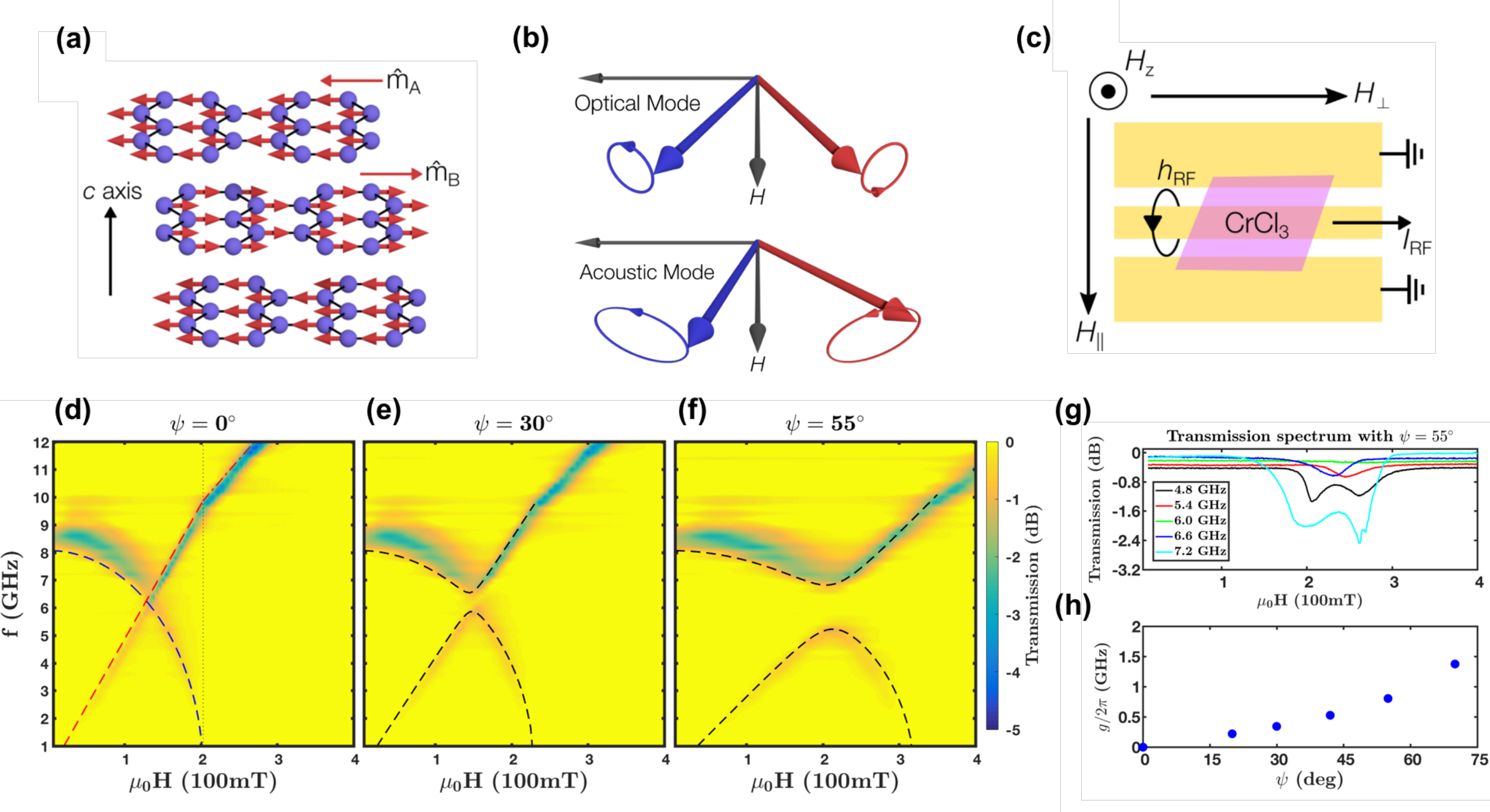}
 \caption{(a) Magnetic structure of bulk CrCl$_3$ below the Ne{\'e}l temperature, and without an applied magnetic field. The blue spheres represent the Cr atoms. The red arrows represent the magnetic moment of each Cr atom with parallel intralayer alignment and antiparallel interlayer alignment. The net magnetization direction alternates between layers, having direction $\hat{m}_{\mathrm{A}}$ ($\hat{m}_{\mathrm{B}}$) on layers in the A (B) magnetic sublattice. (b) Schematic illustrations of the precession orbits for the two sublattice magnetizations in the optical mode and the acoustic mode. (c) Experimental schematic featuring a coplanar waveguide (CPW) with a CrCl$_3$ crystal placed over the signal line. The DC magnetic field is applied along the $H_{||}$ direction. (d)(e)(f) Microwave transmission as a function of frequency and applied field at 1.56 K; the field is applied at an angle of (d) $\psi=0^{\circ}$, (e) $30^{\circ}$, and (f) $55^{\circ}$ from the sample plane. (g) Microwave transmission versus applied field at $\psi=55^{\circ}$ for various frequencies, showing the coupling gap.  (h) The coupling strength $g$ increases with $\psi$, and can be tuned from 0--1.37~GHz. Adapted from Ref.~\onlinecite{MacNeill2019}. \label{CrCl3}}
\end{figure*}

Strong magnon-magnon coupling has been achieved at the interface of two adjacent magnetic layers \cite{chen_strong_2018, klingler_spin-torque_2018}. To realize magnon-magnon coupling within a single material, antiferromagnetic or ferrimagnetic materials with magnetic sublattice structures are required. However, conventional antiferromagnetic resonances lie in THz frequencies which require specialized techniques to probe \cite{bhattacharjee_neel_2018,kampfrath_coherent_2011,baierl_terahertz-driven_2016}. Here, we review the recent observation of magnon-magnon coupling in the layered van der Waals antiferromagnetic (AFM) insulator CrCl$_3$ \cite{MacNeill2019}. Fig.~\ref{CrCl3}(a) illustrates the magnetic structure of CrCl$_3$ below the Ne{\'e}l temperature $T_N \approx$ 14~K, which shows parallel intralayer alignment and antiparallel interlayer alignment of magnetic moments. The alternating magnetization across the layers can be modeled as two sublattice magnetization unit vectors $\hat{m}_A$ and $\hat{m}_B$ in the macrospin approximation. When the external field is applied in the crystal plane, the system is symmetric under twofold rotation around the applied field direction combined with sublattice exchange. This symmetry results in two decoupled modes, the optical mode and the acoustic mode, with even and odd parity under the symmetry, respectively. Fig.~\ref{CrCl3}(b) illustrates the oscillation orbits of the two sublattices magnetization in these two modes. To detect the magnetic resonance, a CrCl$_3$ crystal is placed on a coplanar waveguide (CPW), with crystal $c$ axis normal to the CPW plane and field applied in $H_{||}$ direction, as shown in Fig.~\ref{CrCl3}(c). The microwave transmission signal as a function of frequency and in-plane applied field at $T = 1.56$~K is plotted in Fig.~\ref{CrCl3}(d), which shows an optical mode with finite frequency at zero field, and an acoustic mode with frequency linear in field. The dispersion of the two modes can be understood using the two sublattice Landau-Lifshitz-Gilbert (LLG) equations \cite{MacNeill2019}, whose solutions result in blue and red dashed lines for the optical and the acoustic modes, respectively. By fitting the frequency dispersion, the interlayer exchange coupling field $\mu_0H_{E}=101$~mT and the saturation magnetization $\mu_0M_{s}=409$~mT are obtained, which are consistent with magnetometry measurement results \cite{McGuire22017, MacNeill2019}. A slope change of the acoustic mode dispersion occurs at field $H=2H_E$ because the two sublattices are aligned with the applied field when $H>2H_E$. In this field range, the acoustic mode transforms into uniform ferromagnetic resonance mode, and the Kittel formula is utilized to fit the resonance dispersion. Note that the acoustic and optical modes cross without interaction, and this mode crossing is protected by the two-fold rotation symmetry when field is applied in the crystal plane. In principle, breaking this symmetry can hybridize the two modes and induce an anti-crossing gap, which can be realized by tilting the applied field direction at an angle $\psi$ with respect to the crystal plane. Fig.~\ref{CrCl3}(e) shows the microwave transmission signal with field applied at angle $\psi=30^{\circ}$, which demonstrates a coupling gap generated by the out-of-plane field. The size of the gap increases at angle $\psi=55^{\circ}$, as shown in Fig.~\ref{CrCl3}(f). The modes evolution can be understood as an eigenvalue problem of a two by two matrix derived from the coupled LLG equations \cite{MacNeill2019}, whose solutions result in black dashed lines. The coupling strength $g/2\pi$ is determined as half of the minimal frequency spacing of the modes dispersion, with $g/2\pi\approx0.8$ GHz in the case of $\psi=55^{\circ}$. This is larger than the dissipation rates of the upper and the lower branches, with values $\kappa_{U}/2\pi\approx0.5$ GHz and $\kappa_{L}/2\pi\approx0.2$ GHz respectively, which indicates the realization of strong magnon-magnon coupling at $\psi=55^{\circ}$. The angular dependence of $g$ is shown in Fig.~\ref{CrCl3}(h). By rotating the crystal alignment in an external field, the system can be tuned from a symmetry-protected mode crossing to the strong coupling regime. 

In summary, AFM resonances in CrCl$_3$ have been observed with frequencies within the range of typical microwave electronics (\textless 20 GHz) because of the weak anisotropy and interlayer exchange coupling. This establishes CrCl$_3$ as a convenient platform for studying AFM dynamics. Moreover, strong magnon-magnon coupling within a single material is realized by symmetry breaking induced by a finite out-of-plane field. Because CrCl$_3$ is a van der Waals material which can be cleaved to produce air-stable monolayer thin films \cite{McGuire22017}, these results open up the possibility to realize magnon-magnon coupling in magnetic van der Waals heterostructures by symmetry engineering. \textcolor{black}{While we discuss spin dynamics and mode hybridization of antiferromagnetically-coupled moments in CrCl$_3$, where the exchange interaction is relatively weak due to the interlayer exchange coupling, it is also possible to study magnon modes arising from intralayer exchange coupling within a CrCl$_3$ layer. Due to the much stronger exchange strength, these modes reside in high energy states, out of reach of microwave techniques, yet accessible by optical techniques. This area of development is summarized in Sec. V in this review paper.}

\subsection{Magnon-Magnon Coupling Mechanisms in Layered Systems}
As we have now discussed, the magnon-magnon interaction between acoustic and optical magnons in both SAFs\cite{Sud2020}, and layered van der Waals antiferromagnets\cite{MacNeill2019}, can be controlled by application of a symmetry breaking external field.  There are other avenues to control the coupling between magnons in these types of layered magnets that are being both experimentally and computationally explored\cite{Shiota2020,sklenar2020self}.  To begin this discussion, we first note that the earlier summarized works rely on the generation of spatially uniform optical and acoustic magnons.  In other words, these modes are antiferromagnetic resonances corresponding to a magnon wavenumber near $k = 0$.  To excite finite wavenumber magnons, one strategy is to use micro/nanofabrication tools to lithographically pattern “meandering” antennae directly on top of the magnets\cite{vlaminck2008current,vlaminck2010spin}.  In this way, finite wavenumber magnons can be excited with an in-plane wavevector consistent with the period of the meandering antenna.  For $k \neq 0$ magnons, the dynamic dipolar magnetic field associated with an optical(acoustic) mode can couple the magnon to the other acoustic(optical) magnon\cite{Shiota2020}.  Shiota \textit{et al.}\ realized this dynamic-dipolar coupling of optical and acoustic magnons in SAFs recently in 2020.  For a finite value of $k$, the magnon-magnon coupling was observed to be dependent upon the relative orientation between the in-plane external field and the magnon wavevector.  Additionally, the coupling strength between the optical and acoustic magnons was found to increase monotonically with the wavenumber.  

It is most often the case that when SAFs are fabricated there are only two ferromagnetic layers separated by a nonmagnetic layer.  We now discuss layered magnets with additional layers, so we will differentiate by counting the number of magnetic layers.  With this definition, the majority of measurements in the literature involve bilayers.  Thus,the optical and acoustic magnon modes that are usually reported in SAFs are from bilayers\cite{li2016tunable, wang2018dual,waring2020zero}.  The uniform optical and acoustic antiferromagnetic resonances frequencies of bilayers can modeled and obtained by linearizing a coupled pair of equations that are based on the Landau-Lifshitz-Gilbert (LLG) equation:
\begin{equation}
    \frac{d\mathbf{\hat{m}_A}}{dt} = -\mu_0\gamma \mathbf{\hat{m}_A} \times  [\mathbf{H_{ext}} - H_E\mathbf{\hat{m}_B} -M_s(\mathbf{\hat{m}_A} \cdot \hat{z})\hat{z}],
\end{equation}
\begin{equation}
\frac{d\mathbf{\hat{m}_B}}{dt} = -\mu_0\gamma \mathbf{\hat{m}_B} \times [\mathbf{H_{ext}} - H_E\mathbf{\hat{m}_A} -M_s(\mathbf{\hat{m}_B} \cdot \hat{z})\hat{z}].
\end{equation}
Models based on these coupled equations are called ``macrospin'' models, since each layer is treated as a single large magnetic moment.  Here, $\mathbf{\hat{m}_A}$ and $\mathbf{\hat{m}_B}$ refer to the direction the moment of each layer points.  $\mathbf{H_{ext}}$ and $H_E$ are the external and interlayer exchange fields, and $M_s$ is the magnetization of each layer.  The above macrospin model of a bilayer can be expanded upon.  For example, a tetralayer can be modeled as:
\begin{equation}
    \frac{d\mathbf{\hat{m}_A}}{dt} = -\mu_0\gamma \mathbf{\hat{m}_A} \times  [\mathbf{H_{ext}} - H_E\mathbf{\hat{m}_B} -M_s(\mathbf{\hat{m}_A} \cdot \hat{z})\hat{z}],
\end{equation}
\begin{equation}
\frac{d\mathbf{\hat{m}_B}}{dt} = -\mu_0\gamma \mathbf{\hat{m}_B} \times [\mathbf{H_{ext}} - H_E\mathbf{\hat{m}_A} - H_E\mathbf{\hat{m}_C} -M_s(\mathbf{\hat{m}_B} \cdot \hat{z})\hat{z}],
\end{equation}
\begin{equation}
\frac{d\mathbf{\hat{m}_C}}{dt} = -\mu_0\gamma \mathbf{\hat{m}_C} \times [\mathbf{H_{ext}} - H_E\mathbf{\hat{m}_B} - H_E\mathbf{\hat{m}_D} -M_s(\mathbf{\hat{m}_C} \cdot \hat{z})\hat{z}],
\end{equation}
\begin{equation}
    \frac{d\mathbf{\hat{m}_D}}{dt} = -\mu_0\gamma \mathbf{\hat{m}_D} \times  [\mathbf{H_{ext}} - H_E\mathbf{\hat{m}_C} -M_s(\mathbf{\hat{m}_D} \cdot \hat{z})\hat{z}],
\end{equation}
By solving for the eigenvalues of these four coupled equations, it can be shown that two additional optical and acoustic magnon modes each exist in layered magnets\cite{sklenar2020self}.  We now discuss how this enables ways to control magnon-magnon interactions in SAFs and van der Waals magnets.

\begin{figure}[ht]
\includegraphics[scale = .34]{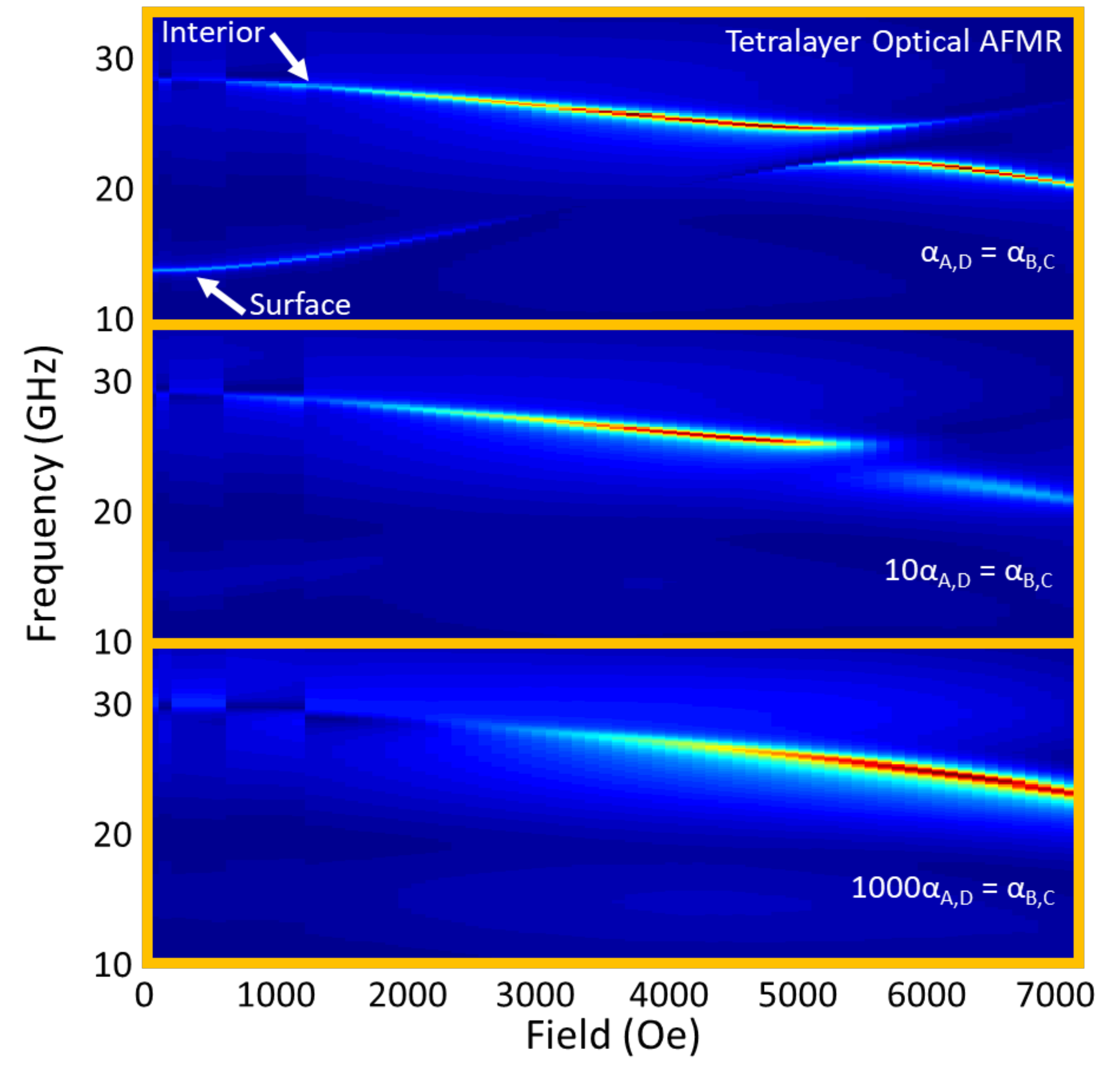}
\caption{The optical mode spectra of a SAF-tetralayer is calculated via micromagnetic simulations in the top panel.  An avoided energy level crossing, indicative of a magnon-magnon interaction between  two optical branches, is observed.  In the middle and bottom panels it is shown that by increasing the damping on the surface layers relative to the interior layers, by a factor of 10 and 1000 respectively, the avoided energy level crossing closes.  In all plots, dark red color indicates a region in field-frequency space where the magnons are strongly excited.  Dark blue indicates regions where no magnons are excited. \label{Tetralayer}}
\end{figure}

It was shown using both a macrospin model and micromagnetic simulations that two optical and two acoustic magnons can be excited in a tetralayer\cite{sklenar2020self}.  The simulations performed used the material parameters for CrCl$_3$, but the results can be generalized to include SAFs.  Each pair of modes resides in a different spatial region of the tetralayer.  For example, a low frequency optical mode can be excited on the surface layers, while a high frequency mode resides on the interior layers.  A magnon-magnon interaction exists between the pair of optical modes or the pair of acoustic modes, even without a symmetry-breaking external field.  Using the optical modes as an example, the optical magnon branch on the surface layers is found to interact with the optical branch of the interior layers due to the exchange field that is generated when magnetization dynamics are present on either surface or interior layers.  This magnon-magnon interaction is evidenced by the avoided energy level crossing present in the optical magnon spectrum shown in Fig.~\ref{Tetralayer}.  \textcolor{black}{Here, we have performed calculations for a SAF-tetralayer assuming that the magnetic layers are permalloy, and that there is an interlayer exchange coupling of $4.14 \times 10^{-14}$A/m.  Because the magnon-magnon interaction is mediated through the dynamic exchange field, there are appealing strategies to control the interaction.}  In a SAF, it is simple to deposit additional ``capping'' layers made of a spin Hall metal like Pt, Ta, or W\cite{hoffmann2013spin}. \textcolor{black}{In fact, current-induced spin-torque excitations of both acoustic and optical modes in SAFs have been demonstrated by Sud et al. very recently} \cite{SudAPL2021}.  By using a DC current bias, a damping like torque can then be applied to the surfaces layers\cite{langenfeld2016exchange}.  This damping like torque effectively can control the strength of the magnon-magnon interaction, by suppressing the dynamic exchange field.  Thus, a strategy to electrically tune the magnon-magnon interaction in layered magnets was proposed\cite{sklenar2020self}. 

To summarize, beyond the use of a symmetry breaking external field, there are other approaches to \textit{tune} the magnon-magnon interaction in layered magnets.  By adjusting the magnon wavenumber, and the in-plane orientation of an external field relative to the wavevector, the dipolar interaction can couple acoustic and optical magnons together\cite{Shiota2020}.  Alternatively, one may consider layered systems with four or more layers.  In this case, by adjusting the damping on the surface layers, relative to the interior layers, magnon-magnon interactions can be controlled\cite{sklenar2020self}.  

\subsection{\textcolor{black}{Exceptional points in Synthetic magnets}}
A more comprehensive overview of exceptional points in magnonic systems takes place in the next section.  Here, we restrict our discussion to SAFs as an outlook for future efforts with these materials.  

In quantum mechanics, non-Hermitian Hamiltonians, invariant under the combination of Parity and Time Reversal operations (PT-symmetric), can harbor exceptional points (EP) in the eigenvalue structure of the Hamiltonian\cite{bender1998real,el2018non}.  To achieve a PT-symmetric Hamiltonian, a system described by the given Hamiltonian should have an equal balance of both gain and loss.  In 2015, it was theoretically proposed that synthetic magnets were ideal macroscopic systems to look for EPs\cite{lee2015macroscopic, galda2016parity}.  Because the LLG equation of motion incorporate damping (loss) through the Gilbert damping parameter, these equations of motion can be thought of as a classical analog to non-Hermitian Hamiltonians, having a complex eigenvalue spectrum.  If two LLG equations are coupled together through an exchange field, it was proposed that spin-torque effects could be used to modify the damping on one magnetic layer relative to another\cite{lee2015macroscopic, galda2016parity}.  In other words (anti)damping-like torques can be used to add (gain)loss to one magnetic layer relative to an adjacent layer.    

Theoretical works calculate the EPs by fixing the values of the damping ratio between layers, and varying the interlayer exchange field\cite{lee2015macroscopic, galda2016parity,yu2020higher,TserkovnyakPRR2020}.  In this way, it can be shown that for a critical exchange field, two complex eigenvalues will bifurcate into two real eigenvalues.  These two real eigenvalues are unique from one another, and correspond to the optical and acoustic magnon modes that have been discussed throughout this section.  In 2019, an experimental breakthrough attempted to mimic this theoretical approach\cite{liu2019observation}.  In Permalloy/Pt/Cobalt magnetic bilayers, the damping ratio was fixed and set by the mismatch between the damping of Permalloy and Cobalt.  By changing the Pt thickness between samples, both the magnitude and sign of the interlayer exchange field, set via the \textcolor{black}{Ruderman–Kittel–Kasuya–Yosida} (RKKY) interaction, was controlled.  By experimentally measuring the optical and acoustic spectra across a series of devices, there was enough sampling of interlayer exchange fields to indicate an exceptional point existed in the Permalloy/Pt/Cobalt structure.

The theoretical and experimental works discussed in the preceding paragraphs clearly demonstrate how the frequencies of the optical and acoustic magnons can be tuned in the vicinity of an EP.  More recent theoretical work suggests that other attributes of magnons, such as non-reciprocal spin wave propagation, can be manipulated in the vicinity of an EP\cite{wang2020steering}.  An unexplored direction to consider involves the magnon-magnon interaction between modes near an EP.  Once an EP is reached, the acoustic and optical branches coalesce, and only one magnon mode is present.  This would appear to preclude a magnon-magnon interaction from existing, as only one mode is present.  The evolution of the magnon-magnon interaction, towards this limit, may offer a new way to control the magnon-magnon interaction through the usage of EPs.   It is also important to note that although theory and experiment tend to focus on bilayers, higher order EPs have been calculated in trilayers\cite{yu2020higher}.  Considering that additional magnon-magnon interactions exist in layered structures beyond bilayers, the potential interplay between higher order EPs and these magnon couplings is vastly unexplored.


\subsection{Non-Hermitian Physics and Exceptional Points}

Non-Hermitian physics, with a focus on open systems where energy conservation does not apply, has been attracting intensive attentions in recent years \cite{2018_NPhys_ElGanainy_nonHermitian,2019_CommnPhys_ElGanainy_nonHermitian,2020_arxiv_Ashida_nonHermitian}. Although non-Hermiticity usually leads to complex eigenvalues and therefore is not favorable when characterizing quantum systems, it has been recognized as the origin of many novel physics that extend beyond quantum mechanics. As a ubiquitous phenomenon, non-Hermiticity has been studied in many different physical realizations based on platforms in electronics \cite{2018_NC_Choi_E,2018_NElectr_Chen_E,2019_PRL_Xiao_E}, microwave \cite{2001_PRL_Dembowski,2016_Nature_Doppler,2020_NPhys_Chen_MW}, acoustics \cite{2016_PRX_Ding,2016_NC_Shi,2019_PRL_Wang_Acoustic}, optics/photonics \cite{2013_NMatt_Feng,2017_Nature_Chen,2017_Nature_Hodaei,2018_Science_Zhou,2018_Nature_Yoon,2019_NMatt_Ozdemir}, optomechanics \cite{2016_Nature_Xu_OM,2018_NPhoton_Zhang_OM} and magnetics \cite{FlebusPRB2020}.

Exceptional point (EP) is one of the most intensively studied novel phenomena in non-Hermitian systems \cite{2012_JPA_Heiss,2014_JPA_Heiss}. An EP, sometimes also referred to as a branch point, is a singularity point on an Riemann surface for a system with two or more coupled modes. It is a special type of degeneracy point, where not only the eigenfrequencies but also the eigenmodes are degenerate. The mathematical singularity at EP is accompanied by a long list of anomalous physical phenomena and applications. At the EP, novel behaviors such as chiral states \cite{2020_NPhys_Wang_Chiral}, unidirectional lasing \cite{2016_PNAS_Peng} and high-sensitivity sensing \cite{2017_Nature_Hodaei,2017_Nature_Chen} can be achieved. The peculiar behaviors can be observed beyond the EP itself. By encircling the EP in the parameter space, asymmetric mode conversion can be achieved \cite{2016_Nature_Doppler,2016_Nature_Xu_OM,2018_Nature_Yoon}. Most importantly, such mode conversions are topologically protected and therefore is insensitive to the specific looping paths as long as the EP is encircled.

In order to observe the EP, a non-Hermitian system has to satisfy the following requirements. First, the system should contain two or more coupled modes. These modes can have the same physical origin ({\em e.g.}, two optical resonances), or they can be different types of modes (one microwave mode and one magnon mode, {\em e.g.}, as will be shown below). Second, the two coupled modes should have identical frequencies. Alternatively, one mode may have a tunable frequency which can vary across the frequency of the other mode. Third, strong coupling should be achievable. An EP can be observed at the phase transition from weak coupling to strong coupling, and therefore strong coupling is a prerequisite for EP. 

Hybrid magnonic systems \cite{HueblPRL2013,ZhangPRL2014,TabuchiPRL2014,2015_PRL_Bai,2014_PRApplied_Goryachev,2016_PRL_Zhang_OMag,2016_PRL_Osada_OMag,2016_PRL_Haigh,2016_SciAdv_Zhang} consist of interacting magnons and at least one other type of information carriers (microwave photons, optical photons, mechanical phonons), and therefore it is natually a non-Hermitian platform, as indicated by the Hamiltonian of the system
\begin{equation}
H= \left( \begin{matrix} \omega_\mathrm{c}-i\kappa_\mathrm{c}/2&g\\ g&\omega_\mathrm{m}-i\kappa_\mathrm{m}/2 \end{matrix}\right),
\label{Eq:Hamiltonian}
\end{equation}
\noindent where $\omega_\mathrm{c}$ and $\kappa_\mathrm{c}$ ($\omega_\mathrm{m}$ and $\kappa_\mathrm{m}$) represent the resonance frequency and dissipation rate of the photon (magnon) mode, respectively, and $g$ is the coupling strength of the magnetic dipole-dipole interaction between magnons and microwave photons. Since the magnon frequency can be easily tuned by an external bias magnetic field, the on-resonance condition can be conveniently achieved. Moreover, when magnons are hybridized with microwave photons, the coupling strength is significantly enhanced by the large spin density in the commonly used magnetic media (YIG, permalloy, etc.) and exceed the dissipation of both the magnon and microwave photon mode, enabling the strong coupling condition. With all the essential requirements satisfied, it is straightforward to observe EPs in a hybrid magnon-microwave photon system.

Exceptional point in hybrid magnonics was first experimentally observed in 2017 \cite{2017_NComm_JQYou}. In this work by Zhang {\em et al.}\ the coupling strength between the magnon mode in a YIG sphere and the microwave resonance in a three-dimensional (3D) copper cavity is adjusted by changing the position of the YIG sphere. This changes the field overlap between the two modes and accordingly affects the coupling strength. Scanning the sphere position induces a phase transition from weak coupling to strong coupling and the unambiguously reveals an EP. In addition, it is also proposed that multiple EPs or high-order EPs \cite{2019_PRB_Zhang_highOrderEP} can be supported on this type of hybrid magnonic systems, which are highly desired for topological mode conversion or EP-based sensing.

There is increasing interests in recent years in pursuing high-dimensional EPs. The concept of synthetic space \cite{2018_Optica_Yuan,2020_Science_Dutt} has been utilized to convert the requirements for higher dimensions to a larger parameter space. However, increased dimensions are accompanied by increased tunability, which is not readily achievable in many physical platforms. This severely limits the experimental demonstration of high dimensional EPs, restricting previous reports to just a handful of investigations on exceptional rings \cite{2015_Nature_Zhen,2017_PRL_Duan,2019_NPhoton_Cerjan} and practically preventing the direct observation of exceptional surfaces (ES) -- a surface formed by a collection of EPs \cite{2019_Optica_Zhen,2019_PRB_Okugawa,2019_PRB_Budich}. Taking advantage of the excellent flexibility in hybrid magnon-microwave photon systems, a four-dimensional synthetic space is constructed in the work by Zhang {\em et al.}, which leads to the first experimental observation of an ES in a non-Hermitian system \cite{2019_PRL_Zhang_ES}. Unique anisotropic behaviors are observed on a special exceptional saddle point on the ES, which can enable multiplexed EP sensing as previously shown in lower dimension systems.

The aforementioned EP observations treat the hybrid magnonic systems as effective parity-time (PT) symmetric systems, with EPs representing the onset of the PT symmetry breaking. Interestingly, anti-PT-symmetry \cite{2016_NP_Peng,2019_LSA_Zhang,2019_Science_Li,2020_PRL_Zhang_synthAntiPT}--the counterpart of PT symmetry--has also been demonstrated in hybrid magnonics \cite{2020_PRL_Yang_antiPT,2020_PRAppl_Zhao_AntiPT}. In these demonstrations, purely imaginary coupling between two interacting modes can be obtained through dissipative coupling, resulting in properties that are conjugate to PT-symmetric systems. In anti-PT-symmetric hybrid magnonic systems, EPs are also observed at the transition from the anti-PT-symmetric phase to the anti-PT-symmetry-broken phase. In addition, such systems allow the observation of other intriguing phenomena such as bound-state-in-continuum (BIC) \cite{2020_PRL_Yang_antiPT}, where the hybrid modes exhibits maximal coherence as well as slow light capability.

Time reversal symmetry breaking is another feature of interest in non-Hermitian systems \cite{2004_EPJD_Harney,2006_JPA_Heiss,2011_PRL_Dietz}. In a hybrid magnonic system, this can be conveniently achieved due to the magnetic nature of magnons. The precessional motion of magnons indicates that they can only coupled with one polarization but not the orthogonal one if microwave photons are circularly polarized. As a result, time reversal symmetry can be broken on a cavity supporting circularly polarized microwave resonances, which can further result in nonreciprocal transmission if the photon polarization is port-dependent \cite{2020_PRAppl_Zhang}. In an alternative approach, time reversal symmetry can be broken using a linearly polarized $\lambda/2$ cavity \cite{2019_PRL_Wang}. Reversing the signal propagation direction changes the phase of the cavity field at the position of the YIG sphere, which in turn induces nonreciprocity in the cavity transmission. Through coupling strength engineering (in the first case) of dissipation engineering (in the second case), tunable nonreciprocity and unidirectional invisibility can be achieved in hybrid magnonic systems.

The study of non-Hermitian physics in hybrid magnonic systems is still in its very early stage. With its remarkable diversity and flexibility, there are enormous opportunities for further exploring novel non-Hermitian physics and applications in hybrid magnonics. For instance, the combination of the large tunability of magnons and their excellent compatibility with other information carriers, together with their rich nonlinearity and capability of breaking time reversal symmetry, may lead to extraordinarily novel behaviors. But along the way there are a number of technical obstacles, such as the limited lifetime and semi-static coupling strength, need to be addressed. The expected outcomes of non-Hermitian hybrid magnonics will not only be primarily in the classical regime, but may also be readily applied to quantum information science.

\section{Directions for quantum sensing and hybrid solid-state quantum systems}
\label{sec:sensing}

\subsection{Introduction to optically addressable solid-state defects}

Atomic defects in semiconductors, such as the negatively charged nitrogen-vacancy (NV$^-$) center in diamond, are a promising platform for quantum technologies due to their long coherence times, atomically defined localization, and optical interface for spin initialization and readout \cite{Gruber1997, Taylor2008, Dolde2011, Heremans2016}. Over the past decade, extensive experimental and theoretical efforts have been devoted to realizing quantum networks \cite{Childress2013,Hensen2015}, registers \cite{Taminiau2014}, and memories \cite{Fuchs2011} in NV-based hybrid quantum systems \cite{MacQuarrie2013,Ovartchaiyapong2014,Barfuss2015,Trifunovic2013,Andrich2017,Kikuchi2017,Wang2020_EControl}. The coupling between solid-state qubits and magnetic materials, in particular, underlies varied applications in quantum information science \cite{Taylor2008, Andrich2017, Degen2017}. In the weak coupling regime, the sensitivity of single spin qubits to local magnetic fields provides a non-invasive, nanoscale probe of magnon physics over a wide range of temperatures \cite{Gruber1997, Taylor2008, Dolde2011, YacobyNComm2015, Heremans2016, Maletinsky2012, Du2017}. As interactions between spin qubits and magnons increase, a hybrid system is formed that enables possible coherent transfer of information between the two systems. In this strong coupling regime, magnon-mediated spin interactions and driving become accessible, providing a future route toward on-chip coherent quantum control \cite{Andrich2017} and \textcolor{black}{entanglement of atomically localized spin qubits} \cite{Trifunovic2013,Fukami2021}. Here we address the properties of solid-state qubits and their interactions with magnetic fields that lend themselves to integration and coupling with magnonic systems for both quantum sensing and coherent control of solid-state spin qubits.

\begin{figure*}[!]
\centering
\includegraphics[width=\textwidth]{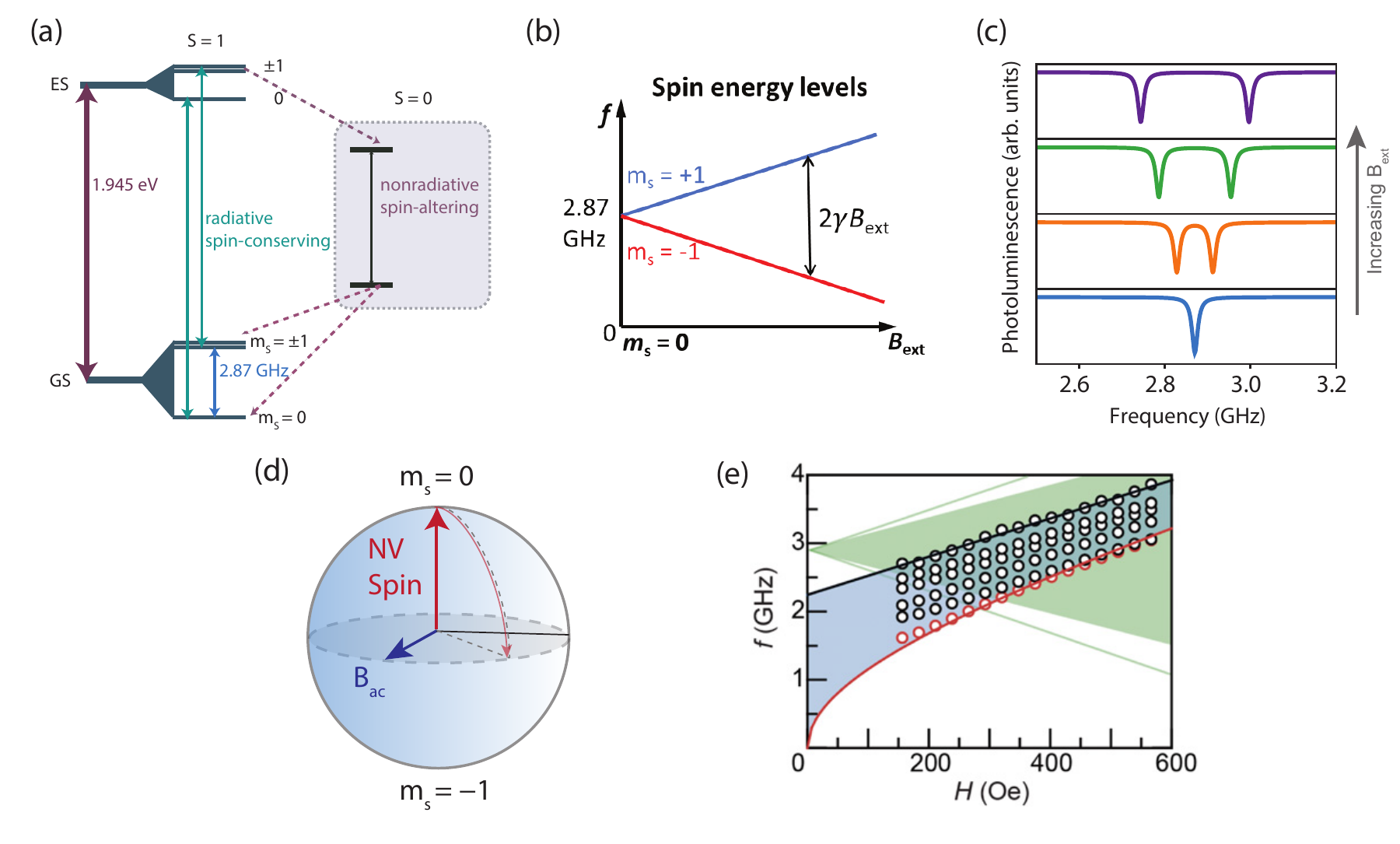}%

\caption{The diamond NV center spin-photon interface and coupling with magnetic fields. (a) The energy level structure showing the spin-dependent transitions of negatively charged NV center, including the spin-conserving radiative transitions and spin-altering nonradiative transitions. Optical excitation preferentially drives the defect into the bright $m_s=0$ state, unless an ac magnetic field on resonance with D is applied to mix the $m_s=\pm1$ and $m_s=0$ sublevels, resulting in reduced photoluminescence. (b) An energy diagram of the NV ESR transitions as a function of external field $B_{ext}$ applied along the NV-axis. In the presence of a magnetic field, the Zeeman effect splits the ground state $m_s=\pm1$ sublevels. (c) Simulated optically detected magnetic resonance (ODMR) spectra for an NV center as a function of magnetic field. The spin-dependent photoluminescence results in two field-dependent dips in the ODMR spectra. (d) Schematic of coherent NV spin rotation in the Bloch sphere. (e) The $m_s=0\leftrightarrow\pm1$ NV transitions (green lines) for different defect orientations can overlap with the magnetostatic surface spin wave dispersion (blue shaded band and black circles) and FMR spin waves (red line and red circles) in a nearby YIG film, enabling strong defect-magnon interactions. Reprinted figure with permission from Ref.~\onlinecite{Kikuchi2017}. Copyright (2017) The Japan Society of Applied Physics.}
\label{NV_fig}
\end{figure*}

To demonstrate the utility of these defect systems as sensors and in magnonic hybrid quantum systems, we will briefly introduce the origin of the spin-photon interface and coupling with nearby magnetic excitations, focusing on the diamond NV$^-$ center as a canonical example. An NV center is formed by a nitrogen atom adjacent to a carbon atom vacancy in one of the nearest neighboring sites of a diamond crystal lattice. The negatively charged NV$^-$ state has a spin triplet ground state with robust quantum coherence and remarkable versatility over a broad temperature range \cite{Rondin2014}. The energy-level structure of the NV$^-$ center and its associated optical transitions are shown in Fig.~\ref{NV_fig}(a). Photons with energy $h\nu \geq 1.945$~eV excite an electron from triplet ground-state (GS) to triplet excited state (ES) in a spin-conserving manner (total-spin $S=1$ and spin-projection quantum numbers $m_s=\pm $1,0). The ES may then decay either through a direct spin-conserving radiative transition to the triplet GS or nonradiatively via a non-spin-conserving intermediate state. This nonradiative transition occurs more readily for sub-levels with $m_s=\pm 1$ than for the $m_s=0$ sub-level. Because the nonradiative transition rates depend on the spin state of the GS, repeating this optical excitation cycle will preferentially drive the system into an $m_s=0$ polarization. This mechanism provides both ~90\% fidelity optical spin initialization and spin-dependent optical readout \cite{Robledo2011, Awschalom2018}.

The GS is split by the diamond crystal field, such that the $m_s=\pm 1$ states are split by $D = 2.87$~GHz from the $m_s=0$ state. Applying a GHz-range ac magnetic field on resonance with $D$ mixes the $m_s=\pm 1$ and $0$ spin projections in the GS manifold and thus enhances the rate for nonradiative decay because there is a proportionally higher $m_s=\pm 1$ population in the excited state. As a result, the photon emission rate decreases, providing a means to optically detect the spin state of the defect and forming the basis for optically detected magnetic resonance (ODMR) measurements. Correlation between the defect’s electron spin state and photon emission rate  in parallel with long coherence times \cite{Togan2010} has enabled both non-invasive initialization and readout of qubits for sensing \cite{Awschalom2018} and long-range entanglement experiments between distant NV centers, a promising step toward the development of quantum communication networks \cite{Hensen2015}. 

Coherent control of the defect spin state can be achieved with ac-magnetic fields (as described above), but also by strain, and ac-electric fields \cite{Whiteley2019a, Miao2019}. Strong interactions with magnetic excitations \cite{Andrich2017} and strain waves \cite{Whiteley2019} in both the host lattice and nearby materials make solid-state quantum defects additionally attractive for applications in hybrid quantum systems where transduction through magnons and phonons occurs at the on-chip level. The addition of their robust spin-photon interface positions quantum defects as prime candidates for bridging the gap between GHz-level qubit excitations and optical photons at the scale of hundreds of THz, which could enable long-distance optical quantum communication. 

In the presence of a local magnetic field applied along the NV axis, the Zeeman effect splits the $m_s=\pm1$ levels in the GS by $2\gamma B$ where the gyromagnetic ratio $\gamma$ is $28$~MHz/mT, as shown in Fig.~\ref{NV_fig}(b). The corresponding (ODMR) spectrum is thereby split into two field-dependent peaks, as demonstrated in Fig.~\ref{NV_fig}(c).  This frequency splitting depends on the orientation of the NV center relative to an external magnetic field \cite{Stanwix2010}, which also sets the coupling with magnonic systems for both hybrid systems and NV-based magnetometers.


For sensing applications the ultimate dc magnetic field sensitivity of NV centers is determined by their electron spin resonance (ESR) linewidths. In addition to dc fields, NV centers also serve as a sensitive probe of ac magnetic fields.  When a microwave field is applied with a frequency matching the spin energy-level-splitting of the NV center, an NV spin will periodically oscillate between two different spin-states in the rotating frame, which is usually referred to as Rabi oscillation \cite{Taminiau2014} as illustrated in Fig.~\ref{NV_fig}(d). The amplitude $B_{ac\perp}$ of the microwave field perpendicular to the NV-axis can be measured by the Rabi oscillation frequency. Any fluctuating magnetic fields at the NV resonance frequency will induce these NV ESR transitions. To date, NV centers have been used to detect weak magnetic fluctuations, e.g. Johnson noise generated by fluctuating electric charges and magnetic noise associated with spin excitations \cite{Du2017,Kolkowitz2015,VanDerSar2015}.

Figure ~\ref{NV_fig}(e) shows the calculated field dependence of magnetostatic surface spin wave modes of a yttrium iron garnet film (contained within the shaded blue region, black and red circles indicate experimentally observed frequencies of surface spin waves and ferromagnetic resonance modes respectively) \cite{Kikuchi2017}. The green lines indicate the field dependence of the NV$^-$ $m_s=0\leftrightarrow\pm1$ transitions for all possible orientations of the defect relative to the external magnetic field \cite{Kikuchi2017}. When the external field is tuned such that the two transitions overlap, the interaction between the NV center and spin waves in the YIG film is maximized. This coupling with magnons can be detected in so-called relaxometry experiments measuring longitudinal spin relaxation, parametrized by $\Gamma_1=1⁄T_1$, the rate at which the nonequilibrium $m_s=\pm1$ and $0$ states relax to their equilibrium distribution. The longitudinal relaxation time $T_1$ is shortened in the presence of magnetic field noise from nearby spin waves at the ESR frequency, thereby affecting the emitted photoluminescence. NV centers exhibit millisecond-long spin relaxation times, enabling field sensitivity down to $10^{-9}$ Tesla to local static and oscillating magnetic fields \cite{Rondin2014}.

\subsection{Quantum sensing of magnons in spintronic systems using NV centers}
In this section, we review the recent progress of NV center based quantum sensing platform and its application to detect magnons in functional spintronic systems. Due to their single-spin sensitivity, NV centers have been demonstrated to be a powerful sensing tool to detect magnetic domains \cite{Tetienne2014,Thiel2019,Velez2019, Broadway2020}, spin transport \cite{Du2017}, and dynamic behaviors \cite{Wolfe2014,Flebus2018,Page2019,McCullian2020a} in a range of emergent magnetic materials. A unique advantage of NV centers results from a combination of the high field sensitivity and nanoscale spatial resolution \cite{Degen2017}. The spatial resolution of an NV center is mainly determined by NV-to-sample distance \cite{Ovartchaiyapong2014}. There are a couple of methods to ensure nanoscale proximity of NV centers to studied materials. The most straightforward way is to deposit materials on a single-crystalline diamond substrate with NV centers implanted a few nanometers below the surface \cite{Bucher2019,Bassett2019}. For certain materials requiring epitaxial growth on specific substrates, patterned diamond nanostructures containing individual NV centers will be used and transferred onto the surface of samples \cite{Wang2020_EControl,Du2017,Burek2012,Lee-Wong2020,Zhang2020}. Thirdly, by employing scanning NV microscopy \cite{Pelliccione2016,Maletinsky2012}, where a micrometer-sized diamond cantilever containing individual NV centers is attached to an atomic force microscope, the NV-to-sample distance can be systematically controlled with nanoscale resolution.

\begin{figure*}[!htb]
\centering
\includegraphics[width=\textwidth]{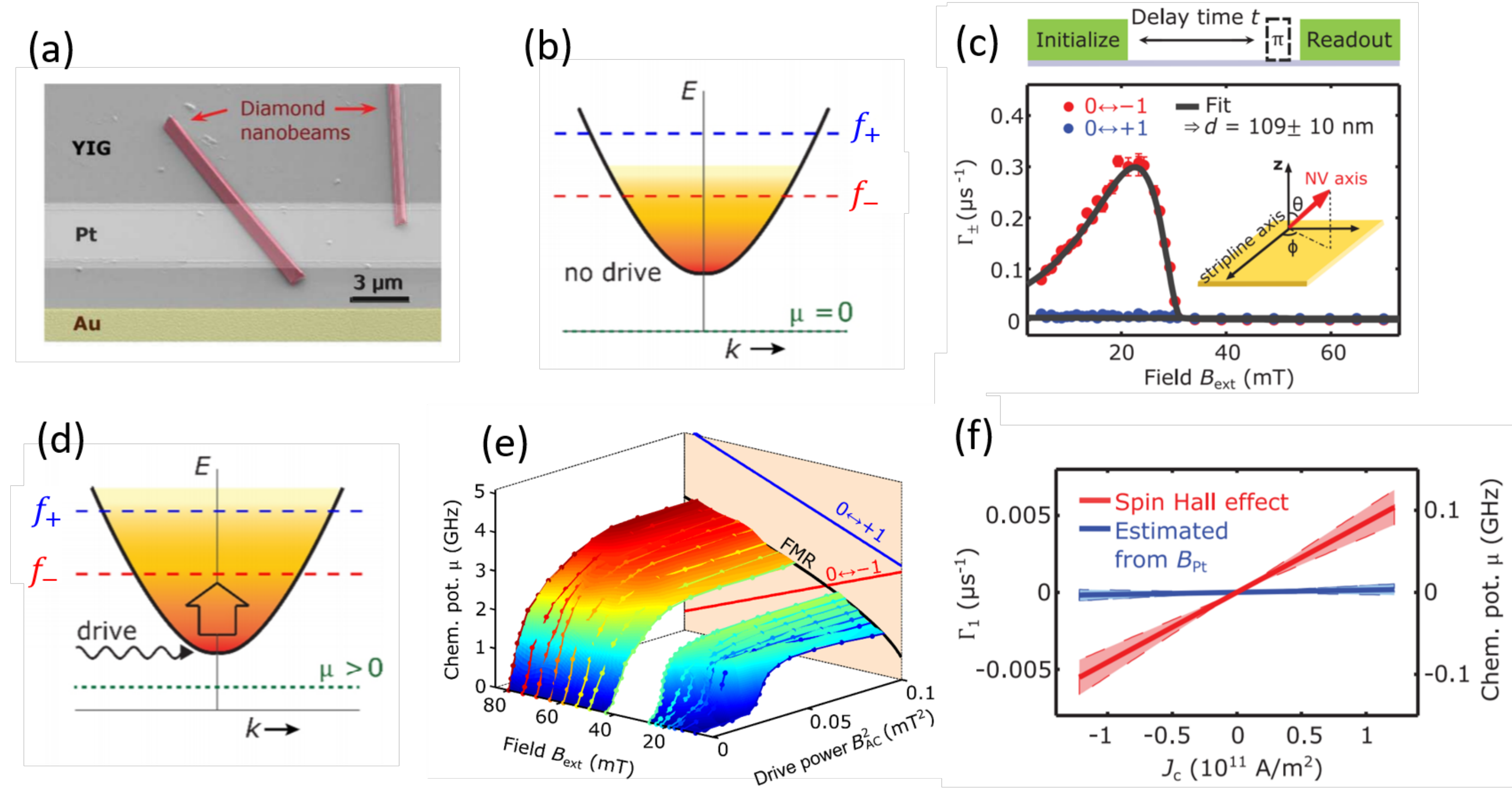}%

\caption{(a) A scanning electron microcopy image showing diamond nanobeams containing individually addressable NV centers positioned on top of a YIG film. A 600-nm-thick Au stripline (false-colored yellow) provides microwave control of the magnon chemical potential in the YIG film and the NV spin states. A 10-nm-thick Pt stripline (false-colored gray) provides spin injection through the spin Hall effect. (b) Sketch of the magnon dispersion and the magnon density with zero chemical potential. NV spin probes the magnon density at the NV ESR frequencies. (c) Field dependence of the measured NV relaxation rate without external spin excitation. (d) Spin excitation effectively increases the magnon chemical potential in YIG. (e) The measured spin chemical potential $\mu$ as a function of microwave power and external field. $\mu$ saturates at the minimum magnon band set by the FMR frequency. (f) The variation of NV spin relaxation rate $\Gamma_1$ and spin chemical potential as a function of the electric current density flowing through the Pt layer. All the figures are taken from Ref.~\onlinecite{Du2017}.}
\label{Sense_fig}
\end{figure*}

Next, we briefly review a recent work using NV centers to detect the spin chemical potential in a magnetic insulator \cite{Du2017, Demokritov2006, Serga2014, Cornelissen2016, Giles2017, Prakash2018, Rezende2018, Olsson2020}. Figure \ref{Sense_fig}(a) illustrates the schematic of the device structure for the measurements, where a 10-nm-thick Pt and a 600-nm-thick Au strip are fabricated on a 20-nm-thick ferrimagnetic insulator Y$_3$Fe$_5$O$_{12}$ film. A diamond nanobeam containing individually addressable NV centers is transferred on the surface of the YIG film. The Au stripline provides microwave control of the NV spin states and the local spin excitations of YIG \cite{Du2017}. The Pt strip provides electrical spin injection through spin Hall effect \cite{Sinova2015,Cornelissen2016}. The NV center is positioned $\sim2$ $\mu$m away from the Au strip and the NV-to-YIG distance is estimated to be $\sim100$ nm. With a moderate external magnetic field applied in the sample plane, the minimum magnon energy band of the YIG film is below the NV ESR frequencies as illustrated in Fig.~\ref{Sense_fig}(b). YIG thermal magnons at the NV ESR frequencies will induce NV relaxation according to the formula: $\Gamma_{\pm}=\frac{k_B T}{f_{\pm} h} \int D(f_{\pm},\mathbf{k})f(\mathbf{k},d)\,d\mathbf{k}$. Here, $f_{\pm}$ and $\Gamma_{\pm}$ are the NV ESR frequencies and relaxation rates of the $m_s = 0 \leftrightarrow \pm1$ transitions, respectively, $T$ is temperature, $D(f_{\pm},\mathbf{k})$ is magnon spectral density, $\mathbf{k}$ is the magnon wave vector and $f(\mathbf{k},d)$ is a transfer function describing the magnon-generated fields at the NV site \cite{Du2017}.

In absence of external spin excitation, NV centers can be used to probe dispersion relationship of the YIG thermal magnons. The top panel of Fig.~\ref{Sense_fig}(c) shows the measurement sequence of the NV relaxometry measurement. A microsecond scale green laser pulse is first applied to initialize the NV spin to the $m_s = 0$ state. During the time delay, YIG magnons at the NV ESR frequencies will couple to the NV spin and accelerate its relaxation. After a delay time $t$, a microwave $\pi$ pulse on the corresponding ESR frequencies is applied to measure the occupation probabilities of the NV spin at the $m_s = 0$, and $m_s = \pm1$ states. The spin-dependent photoluminescence is measured during the green-laser readout pulse. By measuring the integrated photoluminescence intensity as a function of the delay time and fitting the data with a three-level model, NV relaxation rates can be quantitatively obtained. The experimentally measured field dependent NV relaxation rate $\Gamma_-$ shown in the bottom panel of Fig.~\ref{Sense_fig}(c) exhibits a maximum in the region that the corresponding ESR frequency crosses the minimum magnon band of the YIG film, demonstrating the sensitivity of NV centers to noncoherent thermal magnons.

When driving the magnetic system to ferromagnetic resonance (FMR), spin chemical potential will be established in the YIG film [Fig.~\ref{Sense_fig}(d)], leading to an increased magnon density at the NV ESR frequencies and enhanced NV relaxation rates. By measuring the variation of the NV relaxation rates as a function of the driving power, spin wave density and associated spin chemical potential can be directly measured, which is independent of many details of both the quantum sensor and the magnetic material: $\mu=hf_{\pm} (1-\frac{\Gamma_{\pm} (0)}{\Gamma_{\pm}(\mu)})$. The measured magnon chemical potential under resonant condition saturates to the minimum magnon energy band set by the FMR frequency in the large microwave power regime as shown in Fig.~\ref{Sense_fig}(e), demonstrating Bose-Einstein statistics \cite{Du2017,Demokritov2006} of thermal magnons in a magnetic insulator. In the low microwave power regime, the local magnon chemical potential can be systematically controlled by magnetic resonance, in agreement with the theoretical analysis of the underlying multi-magnon processes \cite{Du2017, Flebus2016}.

In addition to magnetic resonance, spin chemical potential could also be electrically established by spin Hall effect \cite{Cornelissen2016, Liu2012, Demidov2017}. When a charge current $J_c$ is applied in the Pt strip, spin currents induced by the spin Hall effect are injected across the YIG/Pt interface, leading to variation of the number of the magnons at the NV ESR frequencies. When the polarization of the spin Hall current is antiparallel to the YIG magnetization, the injected spin current effectively increases the spin chemical potential and the corresponding magnon density at the ESR frequencies. When the polarity of $J_c$ reverses, spin polarization is parallel to the YIG magnetization, leading to a reduced magnon density. This electrically tunable spin chemical potential gives rise to a linear variation of the NV relaxation rate $\Gamma_1 (\Delta\Gamma_-$) on the applied electric current density $J_c$ as shown in Fig.~\ref{Sense_fig}(f). For $J_c = 1.2 \times 10^{11}$~A/m$^2$, $\Gamma_1  (\Delta\Gamma_-)$  is $\sim0.005$ $\mu$s$^{-1}$, which changes sign when the current polarity reverses, in consistent with the spin Hall current injection model \cite{Cornelissen2016}.

\subsection{Magnonic systems for quantum interconnects}
\begin{figure}[!htb]
\centering
\includegraphics[width=3.5in]{ 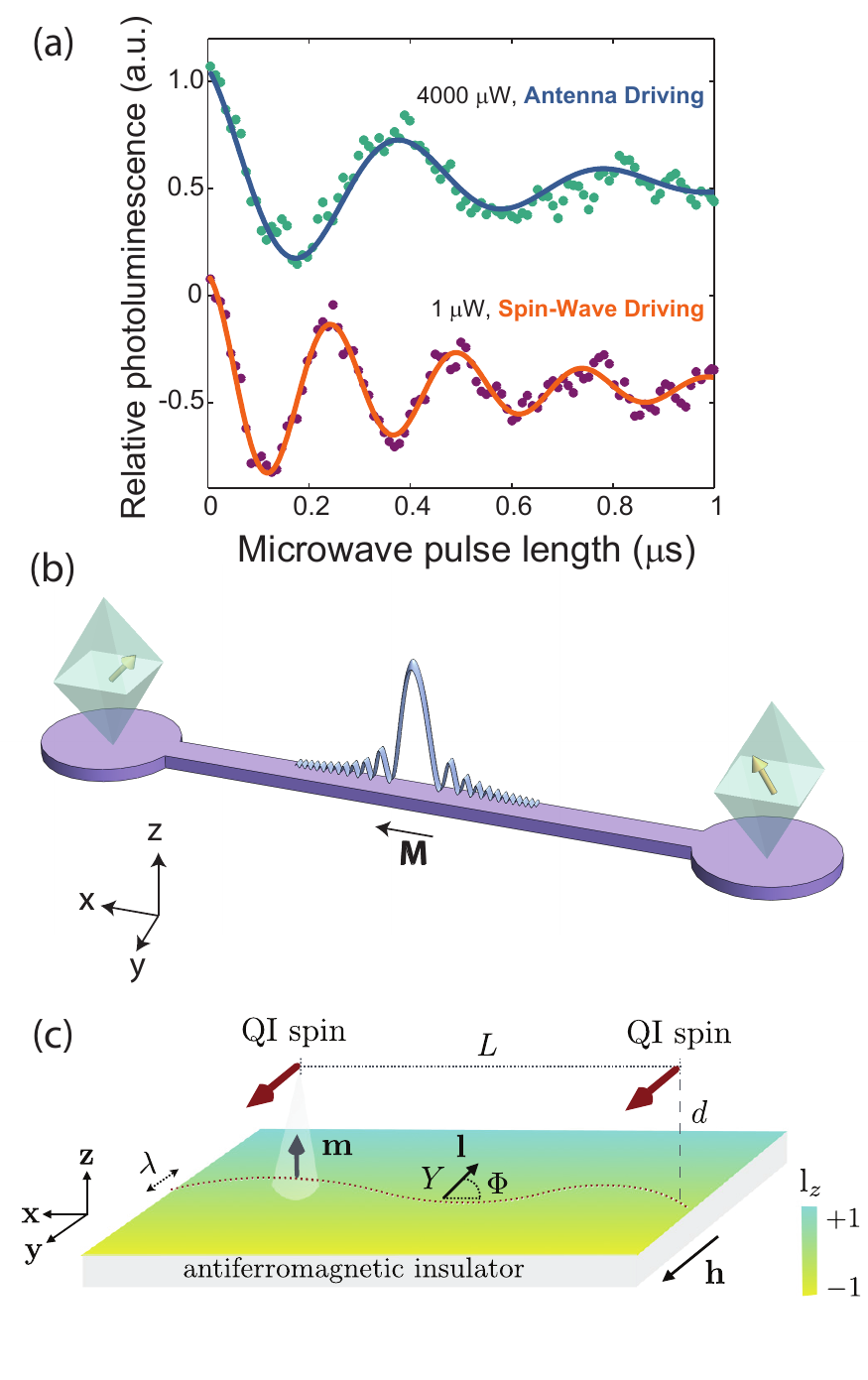}%
\caption{Coupling schemes for distant quantum defect spins mediated by magnons. (a) Microwave driving of surface spin waves in a YIG film coherently drive Rabi oscillations in NV centers over 200~$\mu$m away from the antenna, requiring considerably less power compared to microwave driving of the NV centers coupled through the vacuum. Figure taken from \cite{Andrich2017}. (b) A schematic of the geometry outlined in \cite{Trifunovic2013} where two NV spins are entangled by virtual magnons in a one-dimensional ferromagnet. (c) Quantum defect spins mediated by magnons confined to a domain wall of width $\lambda$ in an antiferromagnet. Reprinted figure with permission from Ref.~\onlinecite{Flebus2019}. {Copyright (2019) by the American Physical Society}.}
\label{remote_NV}
\end{figure}

While quenching of $T_1$ by the broad spectrum of magnetic noise from spin waves is a useful tool for spin wave sensing via noise spectroscopy, it limits the number of coherent operations that can be performed on the coupled qubits.  Placing the lower branch ($m_s=-1$) of the NV transition spectrum in resonance with a magnetostatic surface spin wave mode of a YIG film [Fig.~\ref{NV_fig}(e)], it was demonstrated that directly driving these spin waves could coherently drive NV centers over a distance of several hundred microns up to several millimeters, \cite{Andrich2017,Kikuchi2017,Muhlherr2019}. In particular, it was found that reducing the microwave power reduced noise from off-resonant spin wave excitations and revealed a considerable enhancement of the local microwave driving field mediated by spin waves compared to the driving field directly coupled to the antenna. As observed in Fig.~\ref{remote_NV}(a), this field enhancement can be as high as a factor of several hundred, coherently driving Rabi oscillations in an NV center at a distance of over 200 microns away from the antenna strip line \cite{Andrich2017}. These strong interactions point to magnons in magnetic films as a potential on-chip quantum interconnect between qubits. 

In principle, the magnon-defect interaction is bidirectional. A $\vert0\rangle\leftrightarrow\vert-1\rangle$ transition in the NV center can create or annihilate a magnon in the proximal magnetic film. When two qubits coupled to a magnon waveguide are tuned to be nearly in resonance with long wavelength magnons in the magnet, the two qubits can \textcolor{black}{theoretically be coherently coupled over long distances with the interaction mediated by virtual magnons} \cite{Trifunovic2013,Fukami2021}. This scheme was outlined theoretically considering dipolar interactions with magnons in a one dimensional ferromagnetic waveguide \cite{Trifunovic2013}, as well as along an antiferromagnetic domain wall \cite{Flebus2019}, shown in Fig.~\ref{remote_NV}(b) and (c). It is anticipated that two-qubit gate operations should be achievable over distances on the order of 1 $\mu$m at low ($\le 1$ K) temperatures. Importantly, in this scheme, the qubit splitting must be tuned to be just slightly below the magnon gap of the waveguide so that only virtual magnons are excited. If brought into direct resonance, the interaction with FMR magnons would lead to decoherence.

Controllably tuning the distance and angle between the quantum defect and the magnonic material is important for both sensing applications and coherently coupled interactions. Early studies typically relied on the relatively random placement of defects such as those in drop-cast diamond nanoparticles \cite{Wolfe2014, Labanowski2018}. Advances in deterministic placement of defects, including templated nanoparticle transfer \cite{Andrich2017,Andrich2018, Fukami2019}, nitrogen delta doping of diamond \cite{Ohno2012,Ohno2014,McLellan2016}, local laser annealing \cite{Chen2019}, nano-implantation \cite{Smith2019,Wolfowicz2020}, and placement with scanning probe microscopy techniques \cite{Maletinsky2012,Pelliccione2016} now allow for improved control of qubit positioning. 
 
\subsection{Materials outlook for magnonic hybrid quantum systems}
The materials science challenges of developing optimized hybrid systems present an opportunity to explore emerging spin qubit candidates and magnonic materials. \textcolor{black}{Prior studies of NV center-magnon coupling have focused on YIG owing to its low Gilbert damping coefficient on the order of $10^{-4}$ for thin films \cite{Sun2012,Chang2014, Hauser2016}, resulting in a long magnon spin diffusion length, as well as a convenient overlap of its magnon spectrum with the NV transitions at modest magnetic fields.} YIG films are often grown epitaxially on lattice-matched gadolinium gallium garnet (GGG) \cite{Shone1985,Tang2016}, which is a paramagnet at room temperature, and can present integration challenges in fabrication of diamond/YIG hybrid systems. Further, the magnetic properties of GGG are complex in the millikelvin temperature range relevant to single quasiparticle coherent interactions, resulting in considerable damping \cite{KosenAPLMaterials2019}.

These materials challenges motivate the search for additional low-damping magnetic thin films that can operate as quantum interconnects. Some potential candidates including Co-Fe alloys \cite{Lee2017}, (Ni, Zn, Al) ferrite \cite{Gray2018}, and organic vanadium tetracyanoethylene (V[TCNE]$_x$) \cite{Liu2018,McCullian2020}. Beyond diamond NV centers, developing quantum defects in other host materials \cite{Wolfowicz2020,Weber2010}, opens up opportunities for designing defect-host systems with properties specifically tailored for magnonic hybrid systems. Particularly appealing are defects in semiconductors with mature wafer-scale processing infrastructure, such as silicon carbide (SiC). For instance, excellent optical and spin coherence properties have been demonstrated in divacancy complexes integrated into SiC classical electronic devices \cite{Anderson2019}. \textcolor{black}{Divacancies in SiC have demonstrated smaller zero-field splitting values than diamond NV centers \cite{Koehl2011}, which could allow their ESR transitions to more easily couple to magnetic materials with small magnon gaps at small magnetic fields.} Further afield, qubits consisting of organic molecules containing transition metal ions have recently shown very promising spin and optical properties that can be tuned with ligand chemistry \cite{Bayliss2020}. Coupling such qubits with low-damping organic magnets like V[TCNE]$_x$ could enable all-organic quantum spintronics with qubits separated from magnon waveguides by atomic-scale distances with facile deposition and chemical tunability.  Overcoming material and fabrication challenges both in more traditional systems (e.g., YIG, diamond) and emerging materials is a key priority in the development of integrated magnon/spin-qubit hybrid technologies.

\section{Directions for novel magnonic excitations in quantum materials}
\label{sec:materials}

\subsection{A Brief Background of Magneto-Raman Spectroscopy}
	Magneto-Raman spectroscopy has been shown to be a powerful tool for characterizing collective excitations in magnetic materials \cite{ref1,ref2}. Although the excitations probed are restricted to those with zero total momentum, the optical magneto-Raman scattering process has a decently large scattering cross-section ($\sim 10^{-30}$  cm$^2$ sr$^{-1}$), well-defined symmetry selection rules (dominated by the leading-order electric-dipole approximation), ultra-fine energy resolution ($\sim  0.2$ cm$^{-1}$), and optical diffraction-limited spatial resolution ($\sim$ sub-$\mu$m). As a result, it has made major contributions to the recent blooming fields of two-dimensional (2D) van der Waals (vdW) magnetism and magnetism in systems with spin-orbit-coupling (SOC). In this section of this review article, we will both summarize key recent progress and provide an outlook for optical magneto-Raman spectroscopy applied to these two fields.

 The scattering Hamiltonian by a magnetic system \cite{ref3} is expressed as 
\begin{equation}
{\cal H}'=\sum_{\alpha,\beta} \sum_{\mathbf R} E_1^\alpha E_2^\beta \chi^{\alpha\beta}(\mathbf R),	
\end{equation}
where ${\mathbf E}_1$ and ${\mathbf E}_2$ are the electric fields of the incident and scattered light, respectively, at the magnetic site $\mathbf R$ in the crystal, and $\chi^{\alpha\beta}(\mathbf R)$ is the magnetization-dependent polarizability tensor element at site $\mathbf R$, which can be expanded in terms of the magnetization operators $M_{\mathbf R}$, 
\begin{eqnarray}
\left(\frac{N}{V}\right)\chi^{\alpha\beta}(\mathbf R)
&=& \sum_{\mu} K_{\alpha\beta\mu}(\mathbf R) M_{\mathbf R}^\mu 
+ \sum_{\mu,\nu} G_{\alpha\beta\mu\nu}(\mathbf R)M_{\mathbf R}^\mu M_{\mathbf R}^\nu \notag\\
&&+ \sum_{\mu,\nu,\mathbf r} L_{\alpha\beta\mu\nu} (\mathbf R, \mathbf r)
M_{\mathbf R}^\mu M_{\mathbf R+\mathbf r}^\nu + \cdots	
\label{eq:pol-tensor}
\end{eqnarray}
There are $N$ magnetic sites within the volume of $V$. The first and second terms include operations on the same magnetic site $\mathbf R$ corresponding to single-magnon excitations as well as the elastic scattering off the static magnetic order. The third term contains pairs of operations at adjacent sites $\mathbf R$ and $\mathbf R+ \mathbf r$ representing two-magnon excitations. The tensor forms of coefficients $K_{\alpha\beta\mu}(\mathbf R)$, $G_{\alpha\beta\mu\nu}(\mathbf R)$, and $L_{\alpha\beta\mu\nu} (\mathbf R, \mathbf r)$ are subject to the crystallographic symmetry of the magnetic system \cite{ref3}. 

Below, we will focus on three types of magnetic excitations in 2D vdW and SOC magnetic systems: (a) phonons coupled to the static magnetic order that correspond to time-reversal-symmetry (TRS) broken antisymmetric Raman tensors \cite{ref4,ref5,ref6,ref7,ref8}, (b) single magnon excitations that contain contributions from both the first and second terms in Eq.~\eqref{eq:pol-tensor} \cite{ref9,ref10}, and (c) two-magnon excitations involving pairs of magnons from the Brillouin zone boundaries and center \cite{ref11}. 2D vdW magnetism is an emergent field that started in 2017 with the successful isolation of atomically thin magnetic crystals and the definitive confirmation of ferromagnetism in the monolayer limit \cite{ref12,ref13,BurchNature2018}. While early studies have focused on magnetic ordering in single- and few-layer samples using static magneto-optics \cite{ref12,ref13,ref14,ref15,ref16,ref17,ref18,ref19} (i.e., magneto-optical Kerr effect (MOKE) and magnetic circular dichroism (MCD)) as well as tunneling magnetoresistance \cite{ref20,ref21,ref22,ref23}, optical magneto-Raman spectroscopy has contributed unique insights into the dynamic magnetic excitations in 2D vdW magnets \cite{ref4,ref5,ref6,ref7,ref8,ref9,ref10}. SOC magnetism refers to magnetic systems where SOC is at an energy scale comparable to that of electronic correlations \cite{ref24, ref25}, and it is primarily realized in 5$d$ transition metal oxides (TMOs), such as iridium oxides \cite{ref26, ref27}. While the spin dynamics of 5$d$ TMOs have been primarily probed by hard x-ray scattering \cite{ref28,ref29,ref30,ref31,ref32,ref33,ref34,ref35,ref36} and neutron scattering \cite{ref37,ref38,ref39} (if the sample volume is large enough to overcome the challenges posed by the strong neutron absorption by Ir), optical magneto-Raman spectroscopy has provided complementary information thanks to its superior energy resolution and clean selection rules \cite{ref11, ref40,ref41,ref42,ref43,ref44,ref45}.

\subsection{Example in Phonons Coupled to Static Magnetic Order in a 2D Layered Antiferromagnet}

While ordinary phonons and phonons that are zone-folded by magnetic orders are commonly seen in Raman spectroscopy, they typically correspond to symmetric Raman tensors with TRS \cite{ref2}. The observation of TRS-broken phonon modes with antisymmetric Raman tensors in bulk \cite{ref9} and few-layer CrI$_3$ \cite{ref4,ref5,ref6,ref7,ref8} adds a new variation of phononic excitations, which is now understood as finite-momentum phonons coupled to the static magnetic order \cite{ref5}.

\begin{figure*}
  \includegraphics[width=\textwidth]{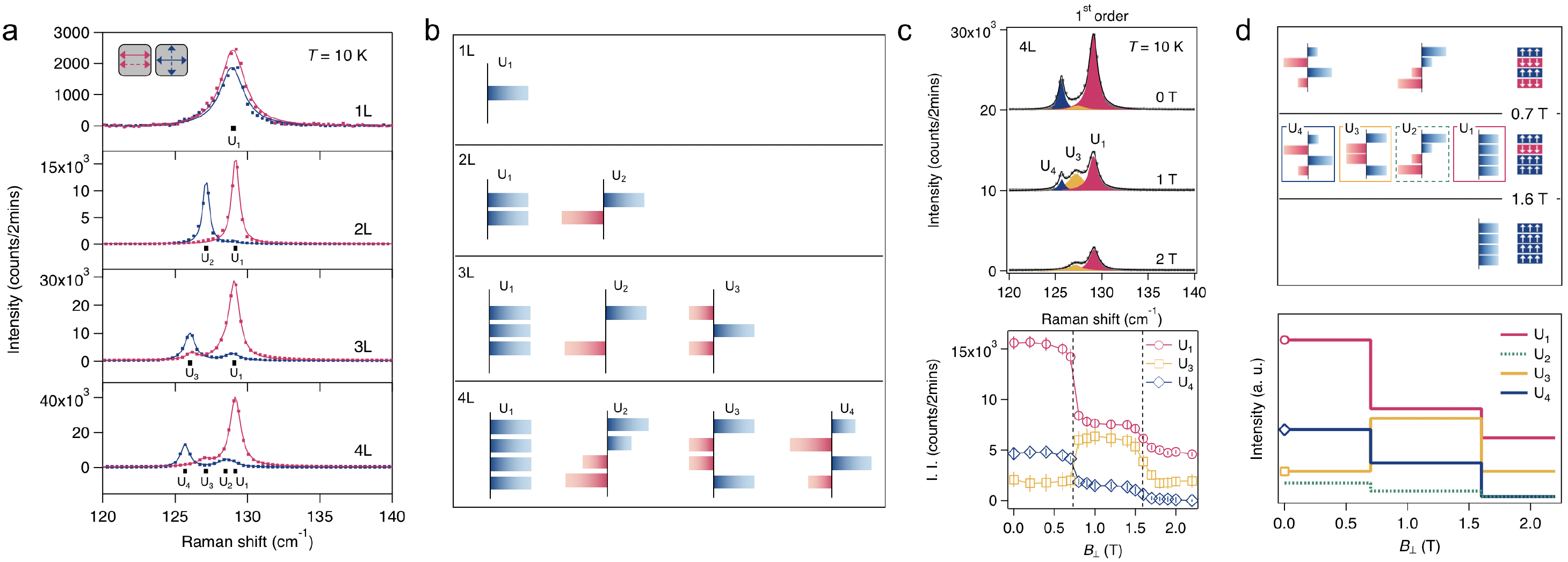}
  \caption{(a) Layer-number-dependent polarized Raman spectra taken on 1-4L CrI$_3$ in both linear parallel (red) and crossed (blue) channels at 10 K. (b) Calculated eigenvectors for the phonon modes in 1-4L CrI$_3$. (c) Magnetic-field-dependent Raman spectra taken on 4L CrI$_3$ in co-circular polarized channel at 10 K. (d) Calculated magnetic field dependence of the four phonon modes of 4L CrI$_3$ in co-circular polarized channel. Figures a, c, and d are adapted from Ref.~\onlinecite{ref5}.}
  \label{fig:Raman}
\end{figure*}

Few-layer CrI$_3$ breaks the out-of-plane translational symmetry, and therefore a singly degenerate phonon mode in monolayer [e.g., A$_{1g}$(D$_{3d}$) modes] gets split into $N$ modes in a $N$-layer CrI$_3$ ($N > 1$) as a result of Davydov splitting [Fig.~\ref{fig:Raman}(a) and (b)]. In the absence of magnetic order, the crystal structure of $N$-layer CrI$_3$ is always centrosymmetric, such that the $N$-split phonon modes have alternating parities under spatial inversion. The highest frequency mode with in-phase atomic displacement between layers is parity-even, and so can be detected in the parallel linear polarization channel. Within the layered antiferromagnetic (AFM) state, the magnetic order of $N$-layer CrI$_3$ is centrosymmetric (parity-even) for odd $N$ and noncentrosymmetric (parity-odd) for even $N$, and so selects parity-even phonons for odd $N$ and parity-odd phonons for even $N$ to restore the centrosymmetry for the product of layered AFM and selected phonons. Such layered-AFM-coupled phonon modes are selected in the crossed polarization channel with a spectral weight proportional to $\vec{U}_i\cdot \vec{M}$, where $\vec{U}_i$ is the eigenvector of the $i^{th}$ phonon mode and $\vec{M}=(1,-1,\cdots,(-1)^{N-1})$ is the axial vector for the layered AFM order of $N$-layer CrI$_3$.

Due to the involvement of layered AFM order in the scattering process for the crossed polarization channel under zero magnetic field, it can be used to track the evolution of the static magnetic order across magnetic phases transitions where $\vec{M}(B)$ changes from layered AFM to ferromagnetic (FM) at $B_c = 0.6$ T for $N = 2$ and $3$ \cite{ref21,ref22}, $B_{c1} = 0.7$ T and $B_{c2} = 1.6$ T for $N \geq 4$ \cite{ref21,ref22}, and $B_c = 2.0$ T for thick flakes and bulk CrI$_3$ \cite{ref9,ref20}. Interestingly, because of the differences in the eigenvector $\vec{U}_i$ among the $N$ phonon modes, individual modes couple differently to the magnetic order and show distinct magnetic field dependencies [Fig.~\ref{fig:Raman}(c) and(d)]. Since the layered magnetism in 2D magnets can also be tuned by external pressure \cite{ref18,ref19}, carrier doping \cite{ref17}, and electric field \cite{ref14,ref15}, the observed magneto-Raman response can be further controlled through device engineering and integrated with spintronic applications. 

The rich magneto-Raman behavior of the layered-magnetism-coupled phonons and their utility in distinguishing different magnetic ordering across layers provide a new avenue to probe magnetic structures in 2D magnets. They also hold promise for 2D magnetism-based spintronic applications.

\subsection{Example in Single-Magnon Excitations in 2D van der Waals Magnets}

Magnons or spin waves are fundamental excitations for magnetically ordered systems. Single-magnon Raman spectroscopy detects the presence of zone-center (zero-momentum) magnons, whereby the Stoke's or anti-Stoke's energy shift is matched with the magnon energy. The selection rule is determined by the first (linear) and/or the second (quadratic) expansions of onsite magnetization operators in Eq.~\eqref{eq:pol-tensor} above. Inelastic x-ray and neutron scattering spectroscopies are widely used in studying magnon excitations in three-dimensional (3D) bulk magnetic materials and can access magnons with momenta across the entire Brillouin zone. Raman spectroscopy, although being restricted to zone-center magnons, is uniquely suitable for probing small-sized samples, such as 2D vdW magnets \cite{ref10}, and further has a finite penetration depth for detecting surface magnetism in 3D magnets \cite{ref9}.

\begin{figure*}
  \includegraphics[width=\textwidth]{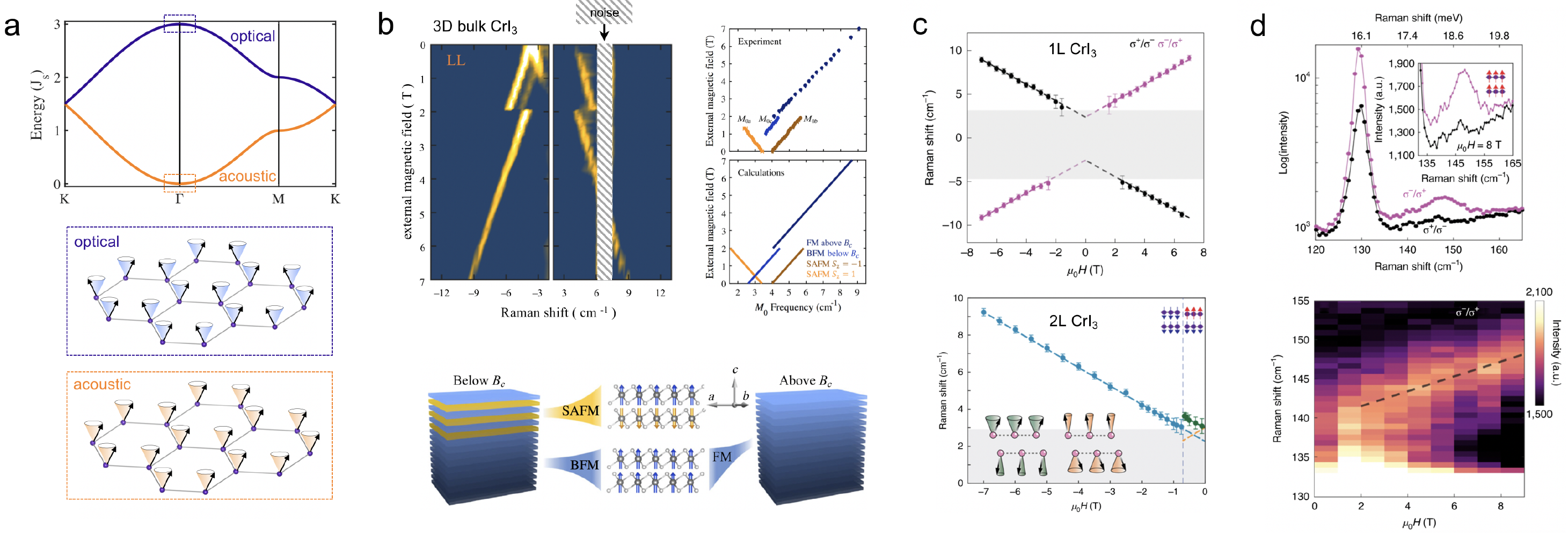}
  \caption{(a) Calculated spin wave dispersion for a honeycomb ferromagnet (i.e., 3D bulk or 1L CrI$_3$), and acoustic and optical spin wave modes. (b) Magnetic-field-dependent acoustic single-magnon Raman spectra taken on 3D bulk CrI$_3$ with the proposed coexisting surface AFM (SAFM) and deep bulk FM (BFM) modes to interpret the experimental data. (c) Magnetic field dependent acoustic single-magnon energy in 1L and 2L CrI$_3$. (d) Magnetic-field-dependent optical single-magnon excitation in 2L CrI$_3$. Figures a, b, and c and d are adapted from Refs. ~\onlinecite{ref4}, \onlinecite{ref9}, and \onlinecite{ref10}, respectively.}
  \label{fig:magnon-dispersion}
\end{figure*}

3D CrI$_3$ was thought to be a ferromagnet as bulk magnetization measurements show a classic FM hysteresis loop \cite{ref46} and inelastic neutron diffraction \cite{ref47} resolves single acoustic and optical magnon branches expected for the honeycomb lattice [Fig.~\ref{fig:magnon-dispersion}(a)]. In contrast to bulk crystals, few-layer CrI$_3$ is revealed to host a layered AFM state through magnetic-field-dependent MOKE and MCD \cite{ref12}, as well as tunneling magnetoresistance \cite{ref20,ref21,ref22,ref23}. The discrepancy between the bulk and few-layer magnetic phases has been reconciled by the observation that the surface layers of bulk CrI$_3$ has layered AFM order, as with few-layer CrI$_3$, while the interior layers show FM order that is consistent with bulk probes. This finding was made by magnetic-field-dependent single-magnon Raman spectroscopy measurements \cite{ref9} [Fig.~\ref{fig:magnon-dispersion}(b)]. Both Stoke’s and anti-Stoke’s Raman spectra successfully capture the single magnon excitations from the acoustic branch in 3D CrI$_3$ and reveal a sudden change at $B_c$ = 2~T that corresponds to the layered AFM to FM transition in thick (i.e., $> 10$~nm) CrI$_3$ flakes. Below $B_c$, three magnon branches are detected in total, a pair with opposite Zeeman shifts for the two spin wave branches with opposite angular momenta in the surface layered AFM and the third branch with a positive Zeeman shift for the spin wave in the interior bulk FM. Above $B_c$, all three branches collapse onto a single mode with a positive Zeeman shift. Both the critical field value and the opposite Zeeman splitting are indicative of the layered AFM state being present in the surface layers of bulk CrI$_3$, which is consistent with magnetic force microscopy measurements on thick CrI$_3$ flakes \cite{ref48}.

Raman spectroscopy is successful in detecting single magnon excitations from the acoustic branch down to the bilayer and monolayer limit, although the signal level is much reduced \cite{ref10} [Fig.~\ref{fig:magnon-dispersion}(c)]. It is of particular interest that the layered AFM order in bilayer CrI$_3$ breaks the spatial inversion symmetry, making the optical spin wave branch Raman-active [Fig.~\ref{fig:magnon-dispersion}(d)]. Compared with the acoustic branch whose frequency at the zone center corresponds to the magnetic anisotropy (i.e., $\propto  J_z-J_x$), the optical branch is at a much higher energy in the terahertz (THz) regime that scales with the average magnetic exchange coupling (i.e., $\propto (J_z+J_x)/2$), and is deemed promising for ultrafast spintronic applications.

In order to achieve a comprehensive understanding of the emerging field of 2D vdW magnetism, it is important to gain insight on the behavior of magnetic excitations. Thus far, Raman spectroscopy has played a crucial role in resolving the fine spectral features that are characteristic of layer-dependent single magnons in vdW magnets.

\subsection{Example in Zone-Center and Zone-Boundary Two-Magnon Excitations in SOC Iridates}

Two-magnon excitations are commonly observed in AFMs where the virtual spin-flip on two adjacent magnetic sites with opposite spins is both permissible and efficient. This process generally happens for the case that the pair of magnons involved are from the Brillouin zone boundaries (i.e., zone-boundary magnons), as illustrated in strongly correlated electron systems like cuprate superconductors \cite{ref49}, vdW AFMs like NiPS$_3$ bulk \cite{ref50} and flakes \cite{ref51}, and traditional bulk magnets like MnF$_2$ \cite{ref52}. 

\begin{figure*} 
  \includegraphics[width=\textwidth]{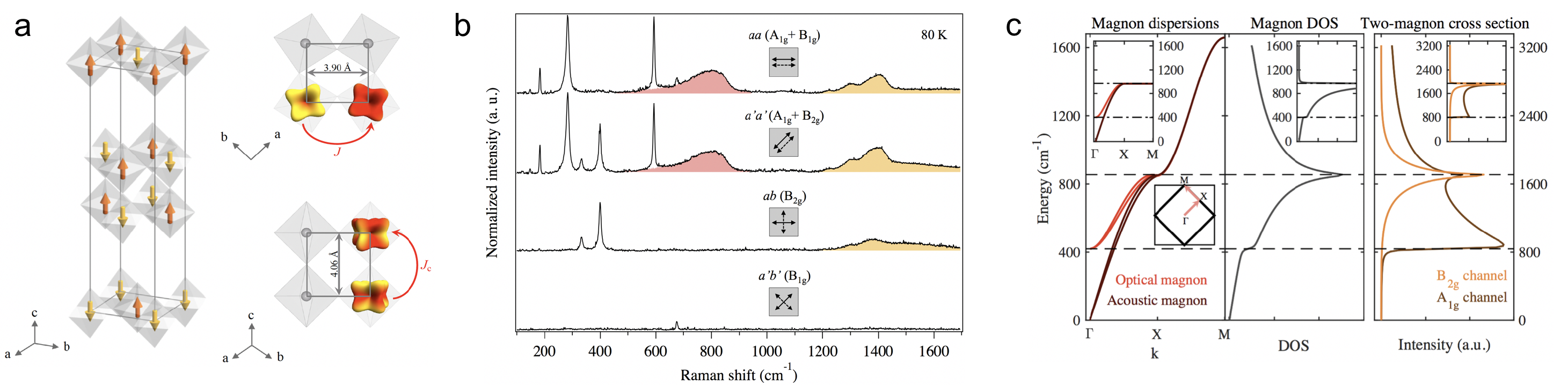}
  \caption{	(a) Illustration of intra and interlayer exchange coupling between $J_{eff} = 1/2$ pseudospins in the bilayer perovskite iridate Sr$_3$Ir$_2$O$_7$. (b) Symmetry-resolved two-magnon excitation Raman spectra taken on Sr$_3$Ir$_2$O$_7$ at 80~K, showing two two-magnon scattering features. (c) Calculated magnon dispersion, magnon density of state (DOS), and two-magnon scattering cross section for a bilayer square-lattice AFM. Figures a--c are adapted from Ref.~\onlinecite{ref11}.}
  \label{fig:iridates}
\end{figure*}

This common scenario of two-magnon excitations requires a revision when there is more than one pair of oppositely aligned spins per magnetic unit cell, for example in chiral AFMs like Mn$_3$X, AFM double-layer perovskite iridates like Sr$_3$Ir$_2$O$_7$, or pyrochlore iridates like R$_2$Ir$_2$O$_7$. Taking double layer perovskite iridate Sr$_3$Ir$_2$O$_7$ as an example, the AFM order happens both within each layer and between the two layers within a bilayer unit, and moreover, the intralayer and interlayer AFM exchange coupling between $J_{eff}=1/2$ pseudospins are at comparable strengths [Fig.~\ref{fig:iridates}(a)]. The latter is in contrast to vdW magnets where intralayer exchange coupling is much stronger than interlayer coupling. In the case of Sr$_3$Ir$_2$O$_7$, the virtual spin flip on the two adjacent sites could be either within the layer or between the two layers within the bilayer. Indeed, this has been shown to be the case in the two-magnon Raman spectra of Sr$_3$Ir$_2$O$_7$ \cite{ref11}.

 Two broad Raman features were detected in the Raman spectra of Sr$_3$Ir$_2$O$_7$, with the one at lower energy ($\sim 800$ cm$^{-1}$) being fully symmetric with respect to the underlying crystal lattice [i.e., A$_{1g}$(D$_{4h}$)] and the other at higher energy ($\sim 1400$ cm$^{-1}$) possessing lower symmetry than that of the lattice [Fig.~\ref{fig:iridates}(b)]. The higher energy feature is also observed at a similar energy and shows the same reduced symmetries in the Raman spectra of the single layer counterpart Sr$_2$IrO$_4$, and therefore is attributed to the zone-boundary two-magnon scattering same as that in Sr$_2$IrO$_4$. The lower energy broad feature, which appears only below the magnetic onset temperature $T_N = 250$ K, is proven of two-magnon nature, but is a new addition in the two-magnon Raman spectra of Sr$_3$Ir$_2$O$_7$ that is not present in Sr$_2$IrO$_4$. Moreover, its full symmetry is unexpected as no magnetic excitations reported so far have shown such high symmetries. From the symmetry perspective, the fully symmetric two-magnon scattering Hamiltonian of Sr$_2$IrO$_4$ shares the same form as (and commutes with) the spin Hamiltonian, and thus no A$_{1g}$ type two-magnon excitations are allowed in Sr$_2$IrO$_4$. In contrast, the A$_{1g}$ two-magnon scattering Hamiltonian of Sr$_3$Ir$_2$O$_7$ differs from its spin Hamiltonian, allowing for the presence of a fully symmetric two-magnon scattering process. From the magnon dispersion perspective, the doubling of the unit cell in Sr$_3$Ir$_2$O$_7$ from Sr$_2$IrO$_4$ doubles the number of spin wave branches, while the interlayer and intralayer exchange coupling being comparable, creates an appreciable energy difference between the acoustic and the optical branch. This fully symmetric two-magnon scattering process therefore involves pairs of magnons from the zone-center optical magnon branch and has an energy nearly twice the optical magnon gap in Sr$_3$Ir$_2$O$_7$ (corresponding to $\sim 400$ cm$^{-1}$ or $\sim  12$ THz) [Fig.~\ref{fig:iridates}(c)].

The symmetry-resolved two-magnon excitations in the bilayer perovskite iridate Sr$_3$Ir$_2$O$_7$ addresses the debate on the nature of its magnetism and sheds light on the mysterious spin wave excitation spectra seen in resonant inelastic x-ray scattering (RIXS). The ultimate potential of well-defined symmetry selection rules in two-magnon scattering is yet to be fully exploited in exploring magnets with strong SOC.

Overall, magneto-Raman spectroscopy has been a powerful experimental technique that is complementary to inelastic x-ray and neutron scattering spectroscopy. It has led to discoveries of new types of magnetic excitations and revealed the nature of novel magnetism in modern magnetic systems. We discuss a few additional possibilities for future applications of magneto-Raman spectroscopy. First, the current lowest detectable Raman shift energy is $\sim 2$ cm$^{-1}$, which is still higher than the spin wave gap energy for many soft magnets (i.e., magnets with small exchange anisotropy), such as XY magnets, magnetic skyrmions, etc. Combining Raman spectroscopy with Brillouin scattering spectroscopy would be one promising avenue to access these low energy magnetic excitations. Second, Raman selection rules allow only for the detection of parity-even modes in centrosymmetric systems, making the high-energy, parity-odd optical spin wave modes often inaccessible by Raman spectroscopy. It may be possible to break the centrosymmetry by applying external parity-odd electrtic fields or by providing controllable non-centrosymmetric defect scattering centers at low symmetry sites in order to access the otherwise forbidden optical magnons. Third, while diffraction-limited magneto-Raman spectroscopy using free-space optics has been primarily used to detect magnetic excitations in low area/volume systems (i.e., 2D film or surface of 3D bulk), tip-enhanced magneto-Raman spectroscopy could be a powerful technique to detect topological edge spin waves locally. Finally, while it is impossible to achieve both high-energy and high-time resolutions, there may be a window to obtain reasonable resolution for both. We propose to access this window by performing time-resolved magneto-Raman spectroscopy for these broad, high-energy two-magnon excitations in Sr$_3$Ir$_2$O$_7$ and to further control the magnetic exchange coupling using light. \textcolor{black}{On the other hand, from a materials science perspective, we also note that the study and understanding of new layered magnets and exotic spin configurations could provide more suitable and interesting candidates to explore the magnon-magnon coupling physics that are addressed in Sec. III. For example, it could be possible to identify other layered magnetic systems at higher temperatures than that of CrCl$_3$. For another example, it could also be possible to identify layered magnetic systems with unconventional magnetic excitations such as Dirac magnons in honeycomb ferromagnets or flat magnon bands in kagome ferromagnets, and explore the magnon-magnon coupling in these systems.}

\subsection{Example in Quantum Molecular Chiral Spin and Magnon Systems}

\textcolor{black}{Designing novel magnetically ordered system at a low-dimensional limit beyond the current 2D layered magnets would offer an alternative route for understanding the 2D magnonic excitation and their rich interactions between microwave photons, optical photons, and phonons \cite{2004_Zutic_1,2015_Hoffmann_2}. The Chirality-Induced Spin Selectivity (CISS) effect, a unique ‘spin filtering’ effect arising from the chirality of low-cost non-magnetic organic materials and their assemblies that lack inversion symmetry, suggests a promising pathway to study the low-dimensional magnetism and their spin wave excitation in a hybrid organic-inorganic quantum system} \cite{2011_Rkikken_4,2012_Naaman_5,2015_Naaman_6,2017_Michaeli_7,2019_Naaman_8}.

\begin{figure*}[htb]
 \centering \includegraphics[width=18 cm]{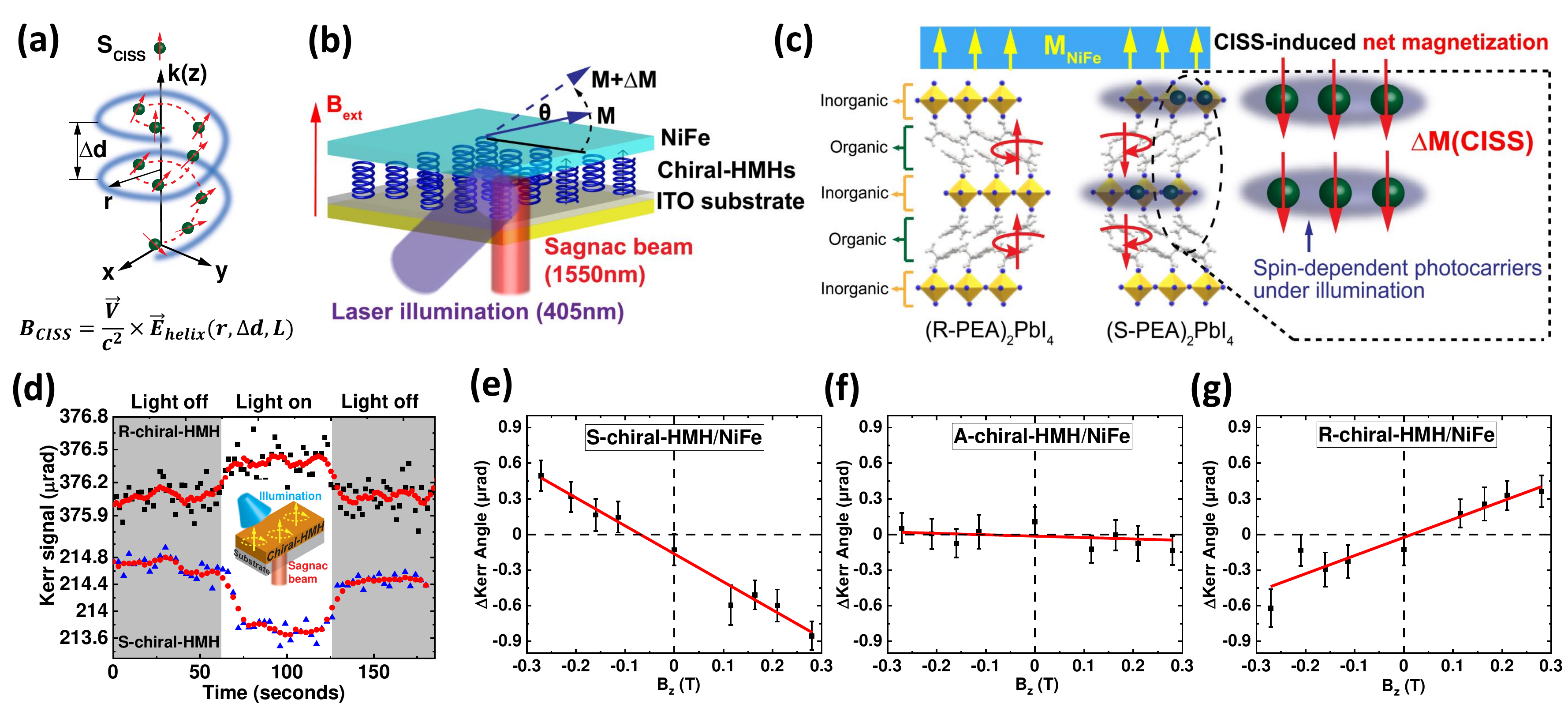}
 \caption{(a) Schematic illustration of the electron (green dots) propagation through a chiral molecular helix and the CISS effect. The red arrows represent the evolution of spin polarization during the transmission. The helical field ($\vec E_{helix}$) induces a magnetic field $\vec B_{CISS}$ and thus alters their spin states. (b) A sketch of light-driven CISS-induced interfacial magnetization in a trilayer structure, illuminated by a 405-nm diode laser from the substrate. The changed magnetic response ($\Delta M$) at the 2D-chiral-HMH/NiFe interface is detected by Sagnac MOKE. (c) Schematic illustration of the photoinduced magnetism at the NiFe/chiral-HMH interface. A net magnetization mediated by spin-dependent photocarriers via the CISS effect is formed in the chiral-HMH structure next to the NiFe layer under the laser illumination. (d) Time trace of measured Kerr (t) signal upon the illumination in the S- and R-chiral-HMH/NiFe heterostructure, respectively. The red line is an adjacent average smoothing of the data \cite{2020_Zhengjie_31}. (e)-(g) show the change in light-induced Kerr angle (black, squares) as a function of external magnetic field. Panels (e), (f), and (g) correspond to S-HMH, A-chiral-HMH and R-HMH respectively. The red line is a linear fit to the data. The Kerr signals have been averaged by four cycles of the illumination to improve the signal-to-noise (Adapted with permission from (\onlinecite{2020_Zhengjie_31}). Copyright (2020) American Chemical Society.)}
 \label{fig:chiral}
\end{figure*}

Chirality is a geometrically distinguishable property of a system that does not possess inversion symmetry, i.e., a mirror plane or glide plane symmetry [Fig.~\ref{fig:chiral}(a)]. Two types of chirality (right- and left-handed, labeled as R and S) can be produced in the molecular structure, as called chiral enantiomers. The CISS effect utilizes chirality to generate a spin-polarized current from chiral (left- or right-handed) enantiomers without the need for magnetic elements \cite{2019_Naaman_8}. The CISS effect has been currently understood as a helicity-dependent spin polarization process via the quantum transport theory \cite{9_1995_Mayer,10_1999_Ray,11_2012_Gutierrez}: The spin polarization is produced by the interplay between the spin moment of an electron ($\vec S_{CISS}$) and an effective magnetic field ($\vec B_{CISS}$) induced by electron propagation through the molecular helix, thereby aligning the electron spin parallel (right-handed helicity) or antiparallel (left-handed helicity) to the direction of the electron’s linear momentum ($\vec S_{CISS} \parallel \pm \vec k$) [Fig.~\ref{fig:chiral}(a)]. This process can be phenomenologically described as \cite{11_2012_Gutierrez,2016_Varela_12,2019_Dalum_13}:

\begin{equation}
\vec S_{CISS} \propto \vec B_{CISS} = \vec v/c^2 \times \vec E_{helix} (r,\Delta d,L)	,
\end{equation}

\noindent where $c$ is the speed of light, $v$ is an electron’s velocity moving through a helical electrostatic field ($\vec E_{helix}$), and $r$, $\Delta d$, and $L$ are the radius, the pitch of the helix, and the number of helical turns, respectively, that determine the magnitude of $\vec E_{helix}$ \cite{11_2012_Gutierrez}. Remarkably, while acting as a magnet with large perpendicular magnetic anisotropy (PMA) to produce an out-of-plane spin current, protected by reduced chiral symmetry, the CISS effect seems not possess an ordering temperature since it does not rely on the itinerant ferromagnetism present in typical magnets, thereby bypassing the challenges associated with thermal fluctuations\cite{1966_Mermin_3}. The robust spin (magnetic) signal it generates can be equivalent to an effective magnetic field of up to tens of Tesla \cite{2019_Naaman_8}. Successful examples of materials exhibiting the CISS effect include quantum dots \cite{2019_Bloom_14,2016_Dor_15}, small molecules \cite{2016_Yang_16,2013_Dor_17}, DNAs \cite{2011_Gohler_18,2011_Xie_19}, and also in 2D chiral hybrid metal halide perovskites (chiral-HMHs) \cite{2018_Long_20,2019_Liu_21}. The 2D chiral-HMHs consist of alternating layers of an inorganic framework of corner-shared lead halide octahedra and chiral organic compounds \cite{2019_Liu_21,2018_Georgieva_22}. The 2D chiral-HMH inherently lacks bulk inversion symmetry and, in principle, is capable of producing an out-of-plane spin current at elevated temperatures via the CISS effect.  	

Here we show the experimental observation of the CISS-induced spin (magnetic) signals in the 2D-chiral-HMH/ferromagnet interface under photoexcitation using an ultrasensitive magneto-optical Kerr effect (MOKE) detection scheme called a Sagnac interferometer \cite{2006_Xia_23}. MOKE has been systematically used for sensing subtle magnetic signals in the semiconductors \cite{2004_Kato_24,2004_Kotissek_25} and metals \cite{2004_Choi_26}, and is particularly suitable for layered 2D magnetism \cite{2017_Gong_27,2017_Huang_28}. Compared to the conventional steady-state MOKE tool that has a Kerr rotation angle resolution of microRadians, a modified fiber Sagnac interferometer is employed based on the design conceived by J. Xia {\em et al.}\ \cite{2006_Xia_23} with dramatically-enhanced sensitivity. The built Sagnac interferometer has a Kerr angle resolution of 50 nanoRadians \cite{2017_McLaughlin_29,2018_McLaughlin_30}. It is therefore reasonable that such a non-invasive spatially resolved Sagnac MOKE approach can locally probe the CISS-induced magnetic signals at the molecular level, providing a direct readout to monitor and to convert light signals into a magnetic response. Figure~\ref{fig:chiral}(b-c) shows a schematic illustration of photoinduced magnetism in the ITO/2D-chiral-HMH/ferromagnet trilayer structure detected by the Sagnac MOKE approach. Under light illumination, photo-excited charge carriers become spin-polarized by the CISS effect as they propagate through the chiral cations in the structure \cite{2016_Dor_15,2013_Dor_17}. The direction of spin polarization is determined by the chirality, resulting in a change of local magnetization at the 2D-chiral-HMH/ferromagnet interface \cite{2020_Zhengjie_31}.

Figure~\ref{fig:chiral}(d) shows the successful observation of light-driven CISS-induced magnetism in the 2D-chiral-HMH/NiFe bilayer detected by Sagnac MOKE. The sample was illuminated at a low laser intensity ($\approx 0.5\ mW$) to suppress the laser-induced heating. By applying a positive out-of-plane magnetic field ($B_z: \pm 210 \  mT$), the Kerr signal in the ITO/(S-Phenylethylamine, PEA)$_2$PbI$_4$/NiFe sample shows a decrease on the order of microRadians ($\Delta\theta_{Kerr}$ $\approx -0.8 \ \mu rad$), supporting our statement that an ultrasensitive detection tool is required to probe this Kerr signal. Whereas there is no similar change found in the A-chiral-HMH sample, the R-chiral-HMH sample exhibits a surprising increase of the Kerr signal under the same illumination. By reversing the magnetic field, the sign of $\Delta\theta_{Kerr}$ is inverted depending on the chirality \cite{2020_Zhengjie_31}, mimicking the nature of magnetization in typical ferromagnets.

The photoinduced magnetism is further corroborated by the magnetic field dependence as shown in Fig.~\ref{fig:chiral}(e)--(g). The change in the Kerr angle $\Delta\theta_{Kerr}$ (B$_z$) exhibits a linear response with the external magnetic field, showing no saturation up to 0.3~T. The sign of the slope depends on the chirality, while there is no field dependence of Kerr signal found in the A-chiral HMH sample. Compared to the Kerr signal obtained in the same NiFe only sample ($\approx 150 \ \mu rad$), the light-driven $\Delta\theta_{Kerr}$ ($\approx 1.0 \ \mu rad$) suggests that the generated interfacial magnetization, $\Delta M$, is around ${0.7\%}$ of the total magnetization of the NiFe layer, corresponding to an effective field of \textcolor{black}{ $\approx \pm 2 \ \mathrm{mT}$}. Here the obtained effective field seems to be orders of magnitude lower than what has been reported in the literature \cite{2011_Gohler_18}, which may be attributed to limited photogenerated carriers under the low light intensity ($ < 0.6 \ mW$) \cite{2013_Dor_17,2016_Yang_16}.

Our work demonstrates the rich characteristics of chiral-HMHs for bridging opto-spintronic applications with the CISS effect \cite{2013_Dor_17,2016_Yang_16,2020_Zhengjie_31}. Magneto-optic Kerr rotation measurements prove that linearly polarized excitation of chiral-HMHs can change the magnetization of an adjacent ferromagnetic substrate. Owing to synthetically tunable optoelectronic properties, the implementation of chiral molecules also offers a novel mutual interconversion between photons, charges, and spins for future ‘opto-magnetism’ applications. The convergence of the CISS-induced large spin polarization, room temperature magnetism, and fascinating optoelectronic/photovoltaic properties of these hybrid layered semiconductors would offer a paradigm shift to transform the field of spintronics using solution-processed hybrid 2D materials, \textcolor{black}{to enable future quantum information technologies.}

\section{Conclusion}

In summary, quantum technologies are promising for the next-generation of computing, sensing, and communication  architectures,  based  on  quantum  coherent transfer and storage of information. It is an emerging field combining fundamental quantum physics, information theory, materials science, and a variety of engineering efforts. For  nascent  quantum  technologies  to  reach  maturity,  a  key  step  is  the  development  of  scalable  quantum  building blocks. Currently, the development of scalable architectures for quantum technologies not only poses challenges in understanding the coupling between disparate quantum systems, but also presents technical and engineering challenges associated with developing chip-scale technologies. 

As a  rapid-growing  subfield  of  quantum  engineering, the developments of hybrid quantum systems have received great attention in recent years. Indeed, the integration  of  different quantum  modules  has  benefited  from  hybrid  quantum  systems,  which  provide  an  important  pathway  for harnessing  the  different  inherent  advantages  of  complementary  quantum  systems,  and  for  engineering  new functionalities. This review article has attempted to summarize and focus on the current frontiers with respect to utilizing magnetic excitations for novel quantum functionality. We have briefly reviewed recent achievements and discoveries in the subject of circuit-based hybrid magnonics systems, layered magnon-magnon systems, quantum-defect sensing of magnons, and novel spin excitations in quantum materials. From each Section, we have attempted to discuss topics spanning the physics fundamentals, technical aspects, and examples of engineered devices and/or material systems.  We have also provided a brief outlook on the future directions of each individual area discussed. 

Overall, magnonics based hybrid systems provide great tunability and flexibility for interacting with various quantum modules for integration in diverse quantum systems. The concomitant rich variety of physics and material selections enable exploration of novel quantum phenomena in materials science and engineering.  The relative ease of generating strong coupling and forming hybrid dynamic system with other excitations also makes hybrid magnonics  a  unique  platform  for  quantum  engineering.

\section*{Acknowledgment}

W.Z. is grateful for the encouragements and support received from his colleagues on the organization and coordination of this perspective article. Work on the manuscript preparation by W.Z. was partially supported by U.S. National Science Foundation under award No. ECCS-1941426. Work on manuscript preparation by D.D.A., F.J.H. and S.E.S. acknowledge support from the U.S. Department of Energy, Office of Science, Basic Energy Sciences, Materials Sciences and Engineering Division. L.R.W. acknowledges support from the University of Chicago/Tohoku University Advanced Institute for Materials Research (AIMR) Joint Research Center. Work on the manuscript preparation by A.H. was supported as part of Quantum Materials for Energy Efficient Neuromorphic Computing, an Energy Frontier Research Center funded by the U.S. DOE, Office of Science, under Award \#DE-SC0019273.  R.H. acknowledges the support by NSF CAREER Grant No. DMR-1760668. C.H.R.D. acknowledges the support from the U.S. National Science Foundation under award ECCS-2029558. J.H. and L.L. acknowledge support from AFOSR under Grant No. FA9550-19-1-0048. D.S. acknowledges the support from U.S. DOE, Office of Science, under Grant No.DE-SC0020992. Work on the manuscript preparation by X.Z. was performed at the Center for Nanoscale Materials, a U.S. Department of Energy Office of Science User Facility, and supported by the U.S. Department of Energy, Office of Science, under Contract No. DE-AC02-06CH11357. We thank Masaya Fukami and Jonathan Karsch for careful reading of the manuscript.

\section*{References}

\bibliography{ref}

\end{document}